\documentclass[12pt]{article}
\usepackage{a4}
\usepackage{array}
\usepackage{calc}

\newlength{\adressabstand}
\input{diagrams}
\newarrow{auf}|---{->}
\newarrow{nach}----{->}
\newarrow{ident}33333
\newarrow{aufsurj}|---{->>}
\newarrow{nachsurj}----{->>}
\newlength{\CDhoehe}                  % Abstand der Zeilen in komm Diags
\newlength{\CDgap}                    % Verringerung des Abstands text -- beginn des CD
\usepackage{ifthen,array}
\newcommand{\ams}{\usepackage{amsfonts,amssymb,amsmath}}%,amsintx}}

\ams
\allowdisplaybreaks[3]
\newlength{\textwidthorig}
\newlength{\oddsidemarginorig}
\newlength{\textheightorig}
\newlength{\topmarginorig}
\def\seitenlaengenabsolut#1 #2 #3 #4 {\setlength{\textwidth}{#1}
                                      \setlength{\oddsidemargin}{#2}
                                      \setlength{\textheight}{#3}
                                      \setlength{\topmargin}{#4}}
\def\seitenlaengenrelzustandard#1 #2 #3 #4 {\setlength{\textwidth}{\textwidthorig+#1}
                                            \setlength{\oddsidemargin}{\oddsidemarginorig+#2}
                                            \setlength{\textheight}{\textheightorig+#3}
                                            \setlength{\topmargin}{\topmarginorig+#4}}
\def\seitenlaengenrelzuvorher#1 #2 #3 #4 {\addtolength{\textwidth}{#1}
                                          \addtolength{\oddsidemargin}{#2}
                                          \addtolength{\textheight}{#3}
                                          \addtolength{\topmargin}{#4}}
\newcommand{\standardseite}{\seitenlaengenrelzuvorher2.2cm -0.8cm 1.8cm -1.5cm }   %3.2%-1.3
\standardseite
\newcommand{\Wegdamit}[1]{}%L"a"st den eingeschlossenen Text unausgedruckt

\newcommand{\nach}{\longrightarrow}      %Abbildungspfeil

\newcommand{\auf}{\longmapsto}           %Abbildungspfeil fuer Elemente
\newcommand{\txtauf}[1]{\auf}            %Abbildungspfeil mit Opt. "Text dr"uber"
\newcommand{\impliz}{\Longrightarrow}    %Implikationspfeil
\newcommand{\invimpliz}{\Longleftarrow}  %Implikationspfeil (umgekehrte Richtung
\newcommand{\gegen}{\rightarrow}         %Konvergenzpfeil
\newcommand{\ident}{\equiv}              %Symbol fuer identisch (3 Striche)
\newcommand{\teilmenge}{\subseteq}       %Symbol fuer Teilmenge
\newcommand{\obermenge}{\supseteq}       %Symbol fuer Teilmenge andersherum
\newcommand{\aeqrel}{\sim}               %Symbol Aequivalenzrelation
\newcommand{\leeremenge}{\emptyset}      %Symbol f"ur leere Menge
\newcommand{\kreuz}{\times}              %Symbol f"ur Kreuz

\newcommand{\einschr}[1]{\mid_{#1}}      %Einschr"ankung auf ...
\newcommand{\dirprod}{\operatorname*{{\text{\Large$\boldsymbol{\kreuz}$}}}}
\newcommand{\betraganpass}[1]%
           {\left| #1 \right|}           %Betragsstriche (variable Gr"o"se) 
\newcommand{\betrag}[1]%
           {\betraganpass{#1}}           %Betragsstriche  
\newcommand{\betragnichtanpass}[1]%
           {\mid #1 \mid}                %Betragsstriche 
\newcommand{\norm}[1]%
           {\parallel #1 \parallel}      %Normstriche  
\newcommand{\erww}[1]%
           {\langle #1 \rangle}          %Erwartungwert-Klammern
\newcommand{\skalprod}[2]%
           {\langle #1,#2 \rangle}       %eckiges Skalarprodukt
\newcommand{\quer}{\overline}            %Strich drueber
\newcommand{\inv}[1]{\frac{1}{#1}}       %Liefert 1/#1 als Bruch
\newcommand{\einhalb}{\inv{2}}           %Liefert 1/2 als Bruch
\newcommand{\im}{\text{im\;}}                          %im als Image (mit platz)
\newcommand{\id}{\:\text{id}}                          %id als ident. Abb.
\newcommand{\inter}{\text{int}\:}                      %int als Inneres 
\newcommand{\elanz}{\#}                                %# f"ur Elementanzahl
\newcommand{\Hom}{\text{Hom}}                          %Hom als Homomorphismengruppe
\newcommand{\field}[1]{\mathbb{#1}}                    %liefert #1 als mathbb-Zeichen
\newcommand{\C}{{\field{C}}}                           %C fuer komplexe Zahlen
\newcommand{\N}{{\field{N}}}                           %N fuer natuerliche Zahlen
\newcommand{\rnkl}[2]{\raisebox{-0.5ex}{$#1$}%
\raisebox{-0.2ex}{{\Large$\setminus$}}\,#2}            %Rechtsnebenklassenfaktorraum
\newcommand{\agb}{{\overline{{\cal A}/{\cal G}}}}      %A/G + Strich
\newcommand{\agbfact}[1][]{\text{$\agb/\!\aeqrel$}}    %A/G\~ + Strich
\newcommand{\Ab}{{\overline{{\cal A}}}}                %A + Strich
\newcommand{\A}{{\cal A}}                              %A ohne Strich
\newcommand{\Gb}{{\overline{{\cal G}}}}                %G + Strich
\newcommand{\AbGb}{{\Ab/\Gb}}                          %A+Strich / G+Strich
\newcommand{\qa}{{\quer{A}}}                           %A als verallg. Zusammenhang 
\newcommand{\holgr}{{\mathbf H}}                       %Holonomiegruppe
\newcommand{\bz}{{\mathbf B}}                          %Basiszentralisator
\newcommand{\Web}{{\text{Web}}}                        %Text "Web"
\newcommand{\Cyl}{\text{Cyl}}                          %Cyl fuer Zylinderfunktionen
\newcommand{\GR}{\Gamma}                               %Graph Gamma
\newcommand{\Ver}{\mathbf{V}}                          %V fuer Vertexmenge
\newcommand{\Edg}{\mathbf{E}}                          %E fuer Kantenmenge
\newcommand{\Pf}{{\cal P}}                             %Menge aller Pfade
\newcommand{\KG}[1]{\Pf_{#1}}                          %Wege_#1
\newcommand{\BB}{\uparrow\uparrow}                     %Relation gleicher Anfangsweg
\newcommand{\EB}{\downarrow\uparrow}                   %Relation End/And-Weg
\newcommand{\BE}{\uparrow\downarrow}                   %Relation Anf/End-Weg
\newcommand{\EE}{\downarrow\downarrow}                 %Relation gleicher Endweg
\newcommand{\notBB}{\mbox{${}\BB{}$\hspace*{-2.8ex}\rule[0.40ex]%
                    {2ex}{0.4pt}\hspace*{0.8ex}}}      %nicht BB
\newcommand{\notEB}{\mbox{${}\EB{}$\hspace*{-2.8ex}\rule[0.40ex]%
                    {2ex}{0.4pt}\hspace*{0.8ex}}}      %nicht EB
\newcommand{\notBE}{\mbox{${}\BE{}$\hspace*{-2.8ex}\rule[0.40ex]%
                    {2ex}{0.4pt}\hspace*{0.8ex}}}      %nicht BE
\newcommand{\notEE}{\mbox{${}\EE{}$\hspace*{-2.8ex}\rule[0.40ex]%
                    {2ex}{0.4pt}\hspace*{0.8ex}}}      %nicht EE
\newcommand{\hyph}{\upsilon}                           %a hyph
\newcommand{\Hyph}{Y}                                  %Y fuer elanz(hyph)
\newcommand{\HyphSet}{{\text{Hyph}}}                   %Text Hyph
\newcommand{\Haar}{{\text{Haar}}}                      %Index Haar
\newcommand{\LG}{{\mathbf{G}}}                         %Liegruppe G (fett)
\newcommand{\Lieg}{{\bf g}}                            %gotisches g (fuer Liegruppe)
\newcommand{\aeqrelzush}[1][]{\sim}                    %Symbol Aequivalenzrelation
\newcommand{\bweg}{e}                                  %spezielle Kante $e$
\newcommand{\nklza}[1][]{\ifthenelse{\equal{#1}{}}     %Z(H_\qa) \ G
                                    {\rnkl{Z(\holgr_\qa)}{\LG}}        
                                   {\rnkl{Z(\holgr_{#1})}{\LG}}}       
\newcommand{\nkla}[1][]{\ifthenelse{\equal{#1}{}}      %B(\qa) \ \Gb
                                    {\rnkl{\bz(\qa)}{\Gb}}        
                                    {\rnkl{\bz(#1)}{\Gb}}}       
\newcommand{\vect}[1]{\vec{#1}}      %Typ
\newcommand{\ListNullAbstaende}{\setlength{\topsep}{0pt}%
                                \setlength{\parskip}{0pt}%
                                \setlength{\partopsep}{0pt}%
                                \setlength{\itemsep}{0pt}%
                                \setlength{\parsep}{0pt}}
\newcommand{\ListNurAnstrichAbstand}{\setlength{\topsep}{0pt}%
                                     \setlength{\parskip}{0pt}%
                                     \setlength{\partopsep}{0pt}%
                                     \setlength{\parsep}{0pt}}
\newenvironment{StandardListe}[2]%
               {\begin{list}%
                      {#1}%
                      {\settowidth{\leftmargin}{M#1}%
                       \settowidth{\labelwidth}{#1}%
                       \settowidth{\labelsep}{M}%
                       #2%
                      }%
                }%
               {\end{list}}%
\newenvironment{EinfachListe}[1]%
               {\begin{StandardListe}{#1}{\ListNullAbstaende}}%
               {\end{StandardListe}}%
               {\begin{StandardListe}{#1}{\ListNurAnstrichAbstand}}%
               {\end{StandardListe}}%
\newcommand{\labelsatz}[1]{#1}
\newcounter{listennr}                      % Wievielte Liste im Dokument
\newlength{\hilfslaenge}
\newlength{\stdlabellaenge}
\newlength{\maximum}
\newcommand{\stdlabel}{}
\newcommand{\Maximum}{}
\newcommand{\iitem}[1][]{\ifthenelse{\equal{#1}{}}%
                           {\item \setlength{\hilfslaenge}{\stdlabellaenge}}%
                           {\item[\labelsatz{#1}\hfill]%
                            \settowidth{\hilfslaenge}{\labelsatz{#1}}}%
                         \ifthenelse{\lengthtest{\maximum < \hilfslaenge}}%
                           {\setlength{\maximum}{\hilfslaenge}%
                            \ifthenelse{\equal{#1}{}}%
                               {\renewcommand{\Maximum}{\stdlabel}}%
                               {\renewcommand{\Maximum}{#1}}}%
                           {}%
                      }      
\makeatletter
\newenvironment{AutoLabelLaengenListe}[2][]%
               {\begin{list}%
                      {\labelsatz{#1}\hfill}%
                      {\stepcounter{listennr}%
                       \settowidth{\leftmargin}{M\labelsatz{\ref{listnr\arabic{listennr}}}}%
                       \settowidth{\labelwidth}{\labelsatz{\ref{listnr\arabic{listennr}}}}%
                       \settowidth{\labelsep}{M}%
                       \settowidth{\stdlabellaenge}{\labelsatz{#1}}%
                       \renewcommand{\stdlabel}{#1}%
                       #2%
                       \renewcommand{\Maximum}{}%
                      }%
                }%
               {\renewcommand{\@currentlabel}{\Maximum}%
                \label{listnr\arabic{listennr}}%
                \end{list}%
                }%
\makeatother
\newenvironment{StandardEinrueckung}[2]%
               {\begin{list}%
                      {#1}%
                      {\settowidth{\leftmargin}{M#1}%
                       \settowidth{\labelwidth}{#1}%
                       \settowidth{\labelsep}{M}%
                       #2%
                      }%
                \item}%
               {\end{list}}%
\newenvironment{Einrueckungpur}[1]%
               {\begin{StandardEinrueckung}{#1}{\ListNullAbstaende}}%
               {\end{StandardEinrueckung}}%
\newenvironment{Einrueckung}[1]%
               {\begin{StandardEinrueckung}{#1}{\setlength{\parsep}{0pt}}}%
               {\end{StandardEinrueckung}}%
\newcommand{\EineZeileGleichung}[2][0.0ex]
           {
            
            \vspace{#1} 
            \noindent
            \hspace*{\fill}
            $\displaystyle{#2}$
            \hspace*{\fill}

            \vspace{#1} 
            
           }

\makeatletter
\newcommand{\EineNumZeileGleichung}[2][0.5ex]
           {
            
            \vspace{#1} 
            \noindent
            \stepcounter{equation}
            \renewcommand{\@currentlabel}{\arabic{equation}}%
            \phantom{(\arabic{equation})}\hspace*{\fill}
            $\displaystyle{#2}$
            \hspace*{\fill}
            (\arabic{equation})

            \vspace{#1} 
            
           }
\makeatother
\newcommand{\breitrel}[1]{\hspace*{\tabcolsep} #1 \hspace*{\tabcolsep}}
\newlength{\abstaug}              %Hilfsl"ange
\newenvironment{AllgUnnumGleichung}[2][1.0ex]%           %#2 gibt die Anordnung der Spalten an
               {
  
                \setlength{\abstaug}{#1}
                \vspace{\abstaug}
                \hspace*{\fill}
                $\begin{array}[t]{#2}
                }%
               {\end{array}$
                \hspace*{\fill}
  
                \vspace{\abstaug}

                }%
               {
  
                \setlength{\abstaug}{#1}
                \vspace{\abstaug}
                $\begin{tabular*}{\textwidth}[t]{#2}
                }%
               {\end{tabular*}$

                \vspace{\abstaug}

               }%
\newcommand{\s}{\\[0ex] }             %Abstand f"ur neue Zeile in Gleichungen   
\newenvironment{StandardUnnumGleichung}[1][0ex]%       %Standard: a = b  (an = ausgerichtet)
               {\renewcommand{\s}{\\[#1] }%
                \begin{AllgUnnumGleichung}{>{\displaystyle}rc>{\displaystyle}l}}%
               {\end{AllgUnnumGleichung}}%
\newcommand{\erl}[1]{\hfill\mbox{\hspace*{1.5em}\small (#1)}}

\newcommand{\erllang}[2][0.5\textwidth]%
              {\hfill\hspace*{1.5em}%
               \begin{minipage}[t]{#1}{\small%
                          \begin{list}{(}{\ListNullAbstaende%
                                          \settowidth{\leftmargin}{(}%
                                          \settowidth{\labelwidth}{(}%
                                          \settowidth{\labelsep}{}%
                                         }%
                          \item#2)%
                          \end{list}}%
               \end{minipage}\\[-0.9ex]
              }%         
\newcommand{\DefBemUmgeb}[1]% %    1. Arg.: Name des Labels, d.h. der Umgebung
           {\newenvironment{#1}[1][]%
                           {\begin{Einrueckung}{{\bf #1}}%
                            \ifx##1\empty\else{{\bf ##1}
                            
                                                        }\fi%
                            }%
                           {\end{Einrueckung}}}
\newcommand{\DefSBemUmgeb}[2]% %    1. Arg.: Name des Labels
           {\newenvironment{#1}[1][]%
                           {\begin{Einrueckung}{{\bf #2}}%
                            \ifx##1\empty\else{{\bf ##1}
                            
                                                        }\fi%
                            }%
                           {\end{Einrueckung}}}
\makeatletter
\newcommand{\DefBspUmgeb}[3]% %    1. Arg.: Name des Labels, d.h. der Umgebung
           {\newcounter{#2}[#3]%
            \newenvironment{#1}[1][]%
                           {\stepcounter{#2}%
                            \renewcommand{\ZaehlerMarke}{\arabic{#2}}%  
                            \renewcommand{\Einzugsname}{{\bf #1 \ZaehlerMarke}}%
                            \begin{Einrueckung}{\Einzugsname}
                            \ifx##1\empty\else{{\bf ##1}\\}\fi%
                            \renewcommand{\@currentlabel}{\ZaehlerMarke}%
                            }%
                           {\end{Einrueckung}}}
\makeatother
\newcommand{\ZaehlerbisEbene}{section}
\newcommand{\Ebenea}{section}
\newcommand{\Ebeneb}{subsection}

\newcommand{\Abschnittnummer}{%
            \ifx\ZaehlerbisEbene\Ebenea{\arabic{section}}%
             \else{%
              \ifx\ZaehlerbisEbene\Ebeneb{\arabic{section}.\arabic{subsection}}%
               \else{\arabic{section}.\arabic{subsection}.\arabic{subsubsection}}%
              \fi}%     
            \fi}     
\newcommand{\Abschnittnummerpunkt}{\Abschnittnummer.}     %  {}%
\newcommand{\Einzugsname}{}
\newcommand{\ZaehlerMarke}{}
\makeatletter
\newcommand{\DefThmUmgeb}[3]% %    1. Arg.: Name des Labels, d.h. der Umgebung
           {\newcounter{#1}[#3]%
            \newenvironment{#1}[1][]%
                           {\stepcounter{#2}%
                            \setcounter{#1}{\value{#2}}%
                            \renewcommand{\ZaehlerMarke}{\Abschnittnummerpunkt\arabic{#1}}%  
                            \renewcommand{\Einzugsname}{{\bf #1 \ZaehlerMarke}}%
                            \begin{Einrueckung}{\Einzugsname}
                            \ifx##1\empty\else{{\bf ##1}
                            
                                                        }\fi%
                            \renewcommand{\@currentlabel}{\ZaehlerMarke}%
                            }%
                           {\end{Einrueckung}}}
\makeatother
\makeatletter
\newcommand{\DefSThmUmgeb}[4]% %    1. Arg.: Name der Umgebung
           {\newcounter{#1}[#3]%
            \newenvironment{#1}[1][]%
                           {\stepcounter{#2}%
                            \setcounter{#1}{\value{#2}}%
                            \renewcommand{\ZaehlerMarke}{\Abschnittnummerpunkt\arabic{#1}}%
                            \renewcommand{\Einzugsname}{{\bf #4 \ZaehlerMarke}}
                            \begin{Einrueckung}{\Einzugsname}
                            \ifx##1\empty\else{{\bf ##1}

                                                        }\fi%
                            \renewcommand{\@currentlabel}{\ZaehlerMarke}%
                            }%
                           {\end{Einrueckung}}}
\makeatother
\newenvironment{Beweis}[1][]%
               {\begin{Einrueckung}{{\bf Beweis}}%
                \ifx#1\empty\else{{\bf #1}

                                            }\fi%
                }%
               {\end{Einrueckung}%
                }%
\newenvironment{Proof}[1][]%
               {\begin{Einrueckung}{{\bf Proof}}%
                \ifx#1\empty\else{{\bf #1}

                                            }\fi%
                }%
               {\end{Einrueckung}%
                }%
               {\begin{Einrueckung}{{\bf \glqq Beweis\grqq}}%
                \ifx#1\empty\else{{\bf #1}
                
                                            }\fi%
                }%
               {\end{Einrueckung}%
                }%
               {\begin{Einrueckung}{{\bf Begr"undung}}%
                \ifx#1\empty\else{{\bf #1}
                
                                            }\fi%
                }%
               {\end{Einrueckung}%
                }%
\newenvironment{Hinrichtung}%
               {\begin{Einrueckungpur}{$\impliz$}}%
               {\end{Einrueckungpur}}%
\newenvironment{Rueckrichtung}%
               {\begin{Einrueckungpur}{$\invimpliz$}}%
               {\end{Einrueckungpur}}%
               {\begin{Einrueckungpur}{\glqq$\teilmenge$\grqq}}%
               {\end{Einrueckungpur}}%
               {\begin{Einrueckungpur}{\glqq$\obermenge$\grqq}}%
               {\end{Einrueckungpur}}%
\newenvironment{SubSet}%
               {\begin{Einrueckungpur}{"$\teilmenge$"}}%
               {\end{Einrueckungpur}}%
\newenvironment{SuperSet}%
               {\begin{Einrueckungpur}{"$\obermenge$"}}%
               {\end{Einrueckungpur}}%
\newcommand{\qed}{\nopagebreak\hspace*{2em}\hspace*{\fill}{\bf qed}}
\newcommand{\ARabic}{\arabic}
\newcommand{\Nummerntypa}{\arabic}   
\newcommand{\Nummerntypb}{\alph}
\newcommand{\Nummerntypc}{\roman}
\newcommand{\Nummerntypd}{\Alph}

\newcommand{\Nra}{\Nummerntypa{Nummera}}            %druckt den Wert des Zaehlers
\newcommand{\Nrb}{\Nummerntypb{Nummerb}}            %entsprechend
\newcommand{\Nrc}{\Nummerntypc{Nummerc}}                
\newcommand{\Nrd}{\Nummerntypd{Nummerd}}                
\newcommand{\ZeichenzuNrTyp}[1]%
           {\ifx#1\ARabic {.}\else{)}%
                  \fi}                              %Numerierung: 1.a)i)A)  
\newcommand{\NrZeicha}{\ZeichenzuNrTyp{\Nummerntypa}}
\newcommand{\NrZeichb}{\ZeichenzuNrTyp{\Nummerntypb}}
\newcommand{\NrZeichc}{\ZeichenzuNrTyp{\Nummerntypc}}
\newcommand{\NrZeichd}{\ZeichenzuNrTyp{\Nummerntypd}}
\newcommand{\ListMarkea}%
           {\Nra\NrZeicha}
\newcommand{\ListMarkeb}%
           {\Nra\NrZeicha\Nrb\NrZeichb}
\newcommand{\ListMarkec}%
           {\Nra\NrZeicha\Nrb\NrZeichb\Nrc\NrZeichc}
\newcommand{\ListMarked}%
           {\Nra\NrZeicha\Nrb\NrZeichb\Nrc\NrZeichc\Nrd\NrZeichd}
\newcommand{\Anfangszeichen}{}
\newcommand{\Anfangspunkt}{}
\newcounter{Schachtelebene}
\newcounter{Hilfszaehler}
\newcommand{\Hilfsbefehl}{}
\newcommand{\Schachtelebene}{\alph{Schachtelebene}}
\makeatletter
\newenvironment{AllgNumerierteListe}[2][]%      %gibt den Wert an, der die max. Laenge besitzt
               {\addtocounter{Schachtelebene}{1}%
		\setcounter{Hilfszaehler}{#2}%
                \renewcommand{\Anfangszeichen}%
                             {\renewcommand{\Hilfsbefehl}{\csname Nummerntyp\Schachtelebene \endcsname}%
                              \Hilfsbefehl{Hilfszaehler}}%
                \renewcommand{\Anfangspunkt}%
                             {\csname NrZeich\Schachtelebene \endcsname}%
                \begin{list}%
                      {\stepcounter{Nummer\Schachtelebene}%
                       \csname Nr\Schachtelebene \endcsname
                       \csname NrZeich\Schachtelebene \endcsname
                       }%
                      {\settowidth{\leftmargin}{M\Anfangszeichen\Anfangspunkt}%
                       \settowidth{\labelwidth}{\Anfangszeichen\Anfangspunkt}%
                       \settowidth{\labelsep}{M}%
                       \setlength{\topsep}{0pt}%
                       \setlength{\parskip}{0pt}%
                       \setlength{\partopsep}{0pt}%
                       \setlength{\itemsep}{0pt}%
                       \setlength{\parsep}{0pt}%
                      }%
                \renewcommand{\@currentlabel}{\csname ListMarke\Schachtelebene \endcsname}%
                }%      
               {\ifthenelse{\equal{}{}}{\setcounter{Nummer\Schachtelebene}{0}}{}
                \addtocounter{Schachtelebene}{-1}%
                \end{list}}
\makeatother
\newenvironment{NumerierteListe}[1]%      %gibt den Wert an, der die max. Laenge besitzt
               {\begin{AllgNumerierteListe}{#1}}
               {\end{AllgNumerierteListe}}
\newenvironment{WeiterNumerierteListe}[1]%      %gibt den Wert an, der die max. Laenge besitzt
               {\begin{AllgNumerierteListe}[Weiter]{#1}}
               {\end{AllgNumerierteListe}}

\newcommand{\UnnumAnfangszeichen}{}
\newcounter{UnnumSchachtelebene}
\newcommand{\UnnumSchachtelebene}{\alph{UnnumSchachtelebene}}
\makeatletter
\newenvironment{UnnumerierteListe}%          
               {\addtocounter{UnnumSchachtelebene}{1}%
                \renewcommand{\UnnumAnfangszeichen}%
                             {\csname UnnumZeich\UnnumSchachtelebene \endcsname}%
                \begin{list}%
                      {\UnnumAnfangszeichen}%
                      {\settowidth{\leftmargin}{M\UnnumAnfangszeichen}%
                       \settowidth{\labelwidth}{\UnnumAnfangszeichen}%
                       \settowidth{\labelsep}{M}%
                       \setlength{\topsep}{0pt}%
                       \setlength{\parskip}{0pt}%
                       \setlength{\partopsep}{0pt}%
                       \setlength{\itemsep}{0pt}%
                       \setlength{\parsep}{0pt}%
                      }%
                }%
               {\addtocounter{UnnumSchachtelebene}{-1}%
                \end{list}}
\makeatother
\newlength{\fktdefhilfslaenge}
\newcommand{\fktdef}[5]%                 %Fast zentriert im Text eingebaut
           {\hspace*{\fill}
            $\begin{array}[t]{cccc}%
            #1: & #2 & \nach & #3 \\    
                & #4 & \auf  & #5
            \end{array}$
            \settowidth{\fktdefhilfslaenge}{$#1$:}
            \hspace*{0.6 \fktdefhilfslaenge}  
            \hspace*{\fill}}
\newcommand{\fktdefpur}[5]%                 %Ohne Abst"ande im Text eingebaut
           {$\begin{array}[t]{cccc}%
            #1: & #2 & \nach & #3 \\    
                & #4 & \auf  & #5
            \end{array}$}
\newcommand{\fktdefabgesetzt}[5]%                %Fast zentriert in zwei extra Zeilen
           {
           
            \hspace*{\fill}
            $\begin{array}[t]{cccc}%
            #1: & #2 & \nach & #3 \\    
                & #4 & \auf  & #5
            \end{array}$
            \settowidth{\fktdefhilfslaenge}{$#1$:}
            \hspace*{0.6 \fktdefhilfslaenge}  
            \hspace*{\fill}
            
            }
\newcommand{\sectioninh}[1]%
           {\section*{#1}%
            \addcontentsline{toc}{section}{#1}}
\newcommand{\anhang}%
           {\appendix
            \sectioninh{Anhang}
            \renewcommand{\Abschnittnummer}{%
                  \ifx\ZaehlerbisEbene\Ebenea{\Alph{section}}%
                  \else{%
                        \ifx\ZaehlerbisEbene\Ebeneb{\Alph{section}.\arabic{subsection}}%
                        \else{\Alph{section}.\arabic{subsection}.\arabic{subsubsection}}%
                        \fi}%     
                  \fi}%
            \renewcommand{\Abschnittnummerpunkt}{\Abschnittnummer.}     
            }            
\newcommand{\anhangengl}%
           {\appendix
            \sectioninh{Appendix}
            \renewcommand{\Abschnittnummer}{%
                  \ifx\ZaehlerbisEbene\Ebenea{\Alph{section}}%
                  \else{%
                        \ifx\ZaehlerbisEbene\Ebeneb{\Alph{section}.\arabic{subsection}}%
                        \else{\Alph{section}.\arabic{subsection}.\arabic{subsubsection}}%
                        \fi}%     
                  \fi}%
            \renewcommand{\Abschnittnummerpunkt}{\Abschnittnummer.}     
            }            
\newlength{\querfhilfsl}              % hilfslaenge f"ur vertauschung der Seitenl"angen (Querformat)

\newlength{\hll}

\newcommand{\zurueck}{\hspace*{-14pt}}
\newcommand{\keinseitenumbr}{\nopagebreak[4]}

\DefThmUmgeb{Theorem}{Theorem}{\ZaehlerbisEbene}
\DefThmUmgeb{Definition}{Definition}{\ZaehlerbisEbene}
\DefThmUmgeb{Satz}{Theorem}{\ZaehlerbisEbene}
\DefThmUmgeb{Proposition}{Theorem}{\ZaehlerbisEbene}
\DefThmUmgeb{Lemma}{Theorem}{\ZaehlerbisEbene}
\DefThmUmgeb{Folgerung}{Theorem}{\ZaehlerbisEbene}
\DefThmUmgeb{Corollary}{Theorem}{\ZaehlerbisEbene}
\DefThmUmgeb{Vorschrift}{Definition}{\ZaehlerbisEbene}
\DefThmUmgeb{Construction}{Definition}{\ZaehlerbisEbene}
\DefSThmUmgeb{FormSatz}{Theorem}{\ZaehlerbisEbene}{\glqq Satz\grqq} 
\DefThmUmgeb{Vermutung}{Theorem}{\ZaehlerbisEbene}
\DefBspUmgeb{Beispiel}{Beispiel}{subsubsection}
\DefBemUmgeb{Bemerkung}
\DefBemUmgeb{Remark}
\DefSBemUmgeb{OffeneFrage}{Offene Frage}
\newcommand{\bdf}{\begin{Definition}}
\newcommand{\edf}{\end{Definition}}
\newcommand{\bvorsch}{\begin{Vorschrift}}
\newcommand{\evorsch}{\end{Vorschrift}}
\newcommand{\bconst}{\begin{Construction}}
\newcommand{\econst}{\end{Construction}}
\newcommand{\bthm}{\begin{Theorem}}
\newcommand{\ethm}{\end{Theorem}}
\newcommand{\bsatz}{\begin{Satz}}
\newcommand{\esatz}{\end{Satz}}
\newcommand{\bprop}{\begin{Proposition}}
\newcommand{\eprop}{\end{Proposition}}
\newcommand{\blem}{\begin{Lemma}}
\newcommand{\elem}{\end{Lemma}}
\newcommand{\bfolg}{\begin{Folgerung}}
\newcommand{\efolg}{\end{Folgerung}}
\newcommand{\bcorr}{\begin{Corollary}}
\newcommand{\ecorr}{\end{Corollary}}
\newcommand{\bbew}{\begin{Beweis}}
\newcommand{\ebew}{\end{Beweis}}
\newcommand{\bpf}{\begin{Proof}}
\newcommand{\epf}{\end{Proof}}
\newcommand{\bwnum}{\begin{WeiterNumerierteListe}}
\newcommand{\ewnum}{\end{WeiterNumerierteListe}}
\newcommand{\bbem}{\begin{Bemerkung}}
\newcommand{\ebem}{\end{Bemerkung}}
\newcommand{\brem}{\begin{Remark}}
\newcommand{\erem}{\end{Remark}}
\newcommand{\bnum}{\begin{NumerierteListe}}
\newcommand{\enum}{\end{NumerierteListe}}
\newcommand{\bunum}{\begin{UnnumerierteListe}}
\newcommand{\eunum}{\end{UnnumerierteListe}}
\newcommand{\bbsp}{\begin{Beispiel}}
\newcommand{\ebsp}{\end{Beispiel}}
\newcommand{\bof}{\begin{OffeneFrage}}
\newcommand{\eof}{\end{OffeneFrage}}
\newcommand{\bgl}{\begin{StandardUnnumGleichung}}
\newcommand{\egl}{\end{StandardUnnumGleichung}}
\newcommand{\zgl}{\EineZeileGleichung}

\newcommand{\znumgl}{\EineNumZeileGleichung}
\newcommand{\berlgl}{\begin{StandardUnnumGleichung}}
\newcommand{\eerlgl}{\end{StandardUnnumGleichung}}
\newcommand{\beinrueck}{\begin{Einrueckungpur}} 
\newcommand{\eeinrueck}{\end{Einrueckungpur}}
\newcommand{\beinflist}{\begin{EinfachListe}} 
\newcommand{\eeinflist}{\end{EinfachListe}}
\newcommand{\beq}{\begin{equation}}
\newcommand{\eeq}{\end{equation}}
\newcommand{\bhin}{\begin{Hinrichtung}}
\newcommand{\ehin}{\end{Hinrichtung}}
\newcommand{\brueck}{\begin{Rueckrichtung}}
\newcommand{\erueck}{\end{Rueckrichtung}}
\newcommand{\bvl}{\begin{AutoLabelLaengenListe}{\ListNullAbstaende}}
\newcommand{\evl}{\end{AutoLabelLaengenListe}}
\newcommand{\df}[1]{{\bf #1}}

\begin{document}
\title{Hyphs and the Ashtekar-Lewandowski Measure}
\author{Christian Fleischhack\thanks{e-mail: 
            Christian.Fleischhack@itp.uni-leipzig.de {\it or}    
            Christian.Fleischhack@mis.mpg.de} \\   
        \\
        \begin{minipage}{0.43\textwidth}
        \begin{center}
        {\normalsize\em Mathematisches Institut}\\[\adressabstand]
        {\normalsize\em Universit\"at Leipzig}\\[\adressabstand]
        {\normalsize\em Augustusplatz 10/11}\\[\adressabstand]
        {\normalsize\em 04109 Leipzig, Germany}\\
        \end{center}
        \end{minipage}
        \begin{minipage}{0.43\textwidth}
        \begin{center}
        {\normalsize\em Institut f\"ur Theoretische Physik}\\[\adressabstand]
        {\normalsize\em Universit\"at Leipzig}\\[\adressabstand]
        {\normalsize\em Augustusplatz 10/11}\\[\adressabstand]
        {\normalsize\em 04109 Leipzig, Germany}\\
        \end{center}
        \end{minipage} \\[-10\adressabstand]
        {\normalsize\em Max-Planck-Institut f\"ur Mathematik in den
                        Naturwissenschaften}\\[\adressabstand]
        {\normalsize\em Inselstra\ss e 22-26}\\[\adressabstand]
        {\normalsize\em 04103 Leipzig, Germany}}
%\date{\today}
\date{January 5, 2000}
\maketitle
\begin{abstract}
Properties of the space $\Ab$ of generalized
connections in the Ashtekar framework are investigated.

First a construction method for new connections is given.
The new parallel transports differ from the original ones only along
paths that pass an initial segment of a fixed path. This is closely 
related to a new notion of path independence. Although we do not restrict
ourselves to the immersive smooth or analytical case, 
any finite set of paths depends on a 
finite set of independent paths, a so-called hyph. This generalizes
the well-known directedness of the set of smooth webs and that of analytical
graphs, respectively.

Due to these propositions, on the one hand, the projections from $\Ab$ to
the lattice gauge theory are surjective and open. 
On the other hand, an induced Haar measure
can be defined for every compact structure group irrespective of the 
used smoothness category for the paths.
\end{abstract}
\newpage
\section{Introduction}
One of the recent approaches to the quantization of gauge 
theories, in particular of gravity, is the investigation of generalized
connections introduced by Ashtekar et al. 
in a series of papers, see, e.g., \cite{a72,a48,a30}. 
Mathematically, there are two main ideas: First, every classical (i.e. 
smooth) connection is uniquely determined by its parallel transports.
These are certain elements of the structure group that are 
in a certain sense smoothly assigned to each path in
the (space-time) manifold and that respect the 
concatenation of paths. Second, quantization here means path
integral quantization. Thus, forget -- as suggested by the Wiener
or Feynman path integral -- the smoothness of the connections being
the configuration variables. Altogether, a generalized connection
is simply defined to be a homomorphism from the groupoid of paths
to the structure group.

At first glance this definition seems to be very rigid. But,  
is there a canonical choice for the groupoid $\Pf$ of paths?
Do we want to restrict ourselves to piecewise analytic or 
immersive smooth paths? When shall two paths be equivalent?
There are lots of "optimal" choices depending on the concrete
problem being under consideration. For instance, for technical reasons
piecewise analyticity is beautiful. In this case it is, in particular, 
impossible that two paths (maps from $[0,1]$ to the manifold $M$)
have infinitely many intersection points provided they do not coincide
along a whole interval. However, since one of the most important
features of gravity is the diffeomorphism invariance, one should 
admit at least smooth paths. Otherwise, a diffeomorphism will no
longer be a map in $\Pf$.
On the other hand, paths that are equal up to the parametrization,
i.e. up to a map between their domains $[0,1]$, should be equivalent.
But, which maps from $[0,1]$ onto itself are reparametrizations?
As well, $\gamma\circ\gamma^{-1}$ are to be equal to the trivial
path in the initial point of the path $\gamma$. This is
suggested by the homomorphy property 
$h_A(\gamma\circ\gamma^{-1}) = h_A(\gamma) h_A(\gamma)^{-1} = e_\LG$
of the parallel transports. What are the other purely algebraic
relations that $h_A$ has to fulfill?

As just indicated, two different definitions are on the market for
a couple of years. Originally, Ashtekar and Lewandowski 
had used the piecewise analyticity \cite{a48}, and later on,
Baez and Sawin \cite{d3} extended their results to the smooth category.
Recently, in a preceding paper \cite{paper2} we considered 
a more general case. At the beginning, 
we only fixed the smoothness category $C^r$, 
$r\in\N^+\cup\{\infty\}\cup\{\omega\}$, and decided whether we 
consider only piecewise
immersed paths or not. Furthermore, we proposed
two definitions for the equivalence of paths.
The first one was -- in a certain sense -- the minimal one:
it identifies $\gamma\circ\gamma^{-1}$ 
with the trivial path as well as reparametrized paths.
The second one identifies in the immersive case paths
that are equal when parametrized w.r.t. the arc length.
The main goal of our paper is a preliminary discussion which results 
are insensitive to the chosen smoothness conditions and which are not.

Foremost, can an induced Haar measure be defined
on the space $\Ab$ of generalized connections in the general 
case? It is well-known that this is indeed possible 
in the analytic case using graphs \cite{a48} and in the smooth case 
using webs \cite{d3}. What common ideas of these cases can be reused
for our problem? 
Looking at the definition
$\Ab_{(r=\omega)}:=\varprojlim_\GR \Ab_\GR$ and 
$\Ab_\Web:=\varprojlim_w \Ab_w$
we see that the label sets $\{\GR\}$ and $\{w\}$ 
of the projective limit are in both cases not only 
projective systems, but also directed systems. This means that, e.g., for
every two graphs there is a third graph such that every
path in one of the first two graphs is a product of paths (or their
inverses) in the third graph. The analogous result holds for the webs.
In the analytical case this can be seen very easily \cite{a48}, for
the smooth one we refer to the paper \cite{d3} by Baez and Sawin.
In \cite{paper2} we defined $\Ab$ in general by
$\Ab_{(r)}:=\varprojlim_\GR \Ab_\GR$ whereas, of course, here the graphs
are in the smoothness category $C^r$. This definition has the drawback
that the projective label set $\{\GR\}$ is no longer directed.
But, nevertheless, note that we have shown \cite{paper2}
in the immersive smooth category that 
$\varprojlim_w \Ab_w$ and $\Ab_{(\infty)}=\varprojlim_\GR \Ab_\GR$ 
are homeomorphic.
Hence that we can hope to find another appropriate label set
for the case of arbitrary smoothness that generalizes the notion of webs
and that gives a definition of the space of generalized 
connections which is equivalent to that using graphs.

In the first step we will investigate a condition for the independence
of paths. When can one assign parallel transports to paths  
independently? As we will see, a finite set $\{\gamma_i\}$ of paths 
is already independent
when every path $\gamma_i$ contains a point $v_i$ such that one of
the subpaths of $\gamma_i$ starting in $v_i$ is 
non-equivalent to every subpath of the $\gamma_j$ with $j<i$.
Sets of paths fulfilling this condition will be called hyph.
Obviously, the edges of a graph are a hyph as well as the curves
of a web. The crucial point is now: For every two hyphs there is
a hyph containing them. In other words, the set of hyphs is 
directed as the set of graphs ($r=\omega$) and that of webs $(r=\infty)$.
This ensures the existence of an induced Haar measure in $\Ab_{(r)}$
for arbitrary $r$.
Moreover, as a by-product we get an explicit construction for 
connections that differ from a given one only along paths that
are not independent of an arbitrary, but fixed path. This immediately
leads to the surjectivity of the projections $\pi_\GR$ from the continuum
to the lattice theory as well as that of $\pi_w$ and $\pi_\hyph$ 
projecting to the webs and hyphs, respectively. 
Furthermore, we prove that $\pi_\GR$ is open. 
In Section \ref{sect:AL-measure} we extend the definition of the 
Ashtekar-Lewandowski measure to arbitrary smoothness categories.
Finally, we discuss in which cases the regular connections form a
dense subset in $\Ab_{(r)}$.

\section{Notations}
In this section we shall recall the basic definitions and 
notations introduced in \cite{paper2}. For further, detailed information
we refer the reader to that article.

Let there be given a finite-, but at least two-dimensional manifold $M$ 
and a (not necessarily compact) Lie group $\LG$.
Furthermore we fix an $r\in\N^+\cup\{\infty\}\cup\{\omega\}$ and decide
whether we work in the category of piecewise immersive maps or not.
In the following we will usually say simply $C^r$ referring to these choices.

A path is a piecewise $C^r$-map from $[0,1]$ to the manifold $M$. 
A graph consists of finitely many non-self-intersecting edges 
whose interiors are disjoint and contain no vertex.
Paths in graphs are called simple, and finite products of simple 
paths are called finite paths.
Two finite paths are equivalent if they coincide up to piecewise
$C^r$-reparametrizations
or cancelling or inserting retracings $\delta\circ\delta^{-1}$.
The set of (equivalence classes of) finite paths is denoted by $\Pf$.
In what follows, we say simply "path" instead of "finite path" and
simply "graph" instead of "connected graph".

A generalized connection $\qa\in\Ab$ is a homomorphism
$h_\qa:\Pf\nach\LG$. For every graph with edges $e_i\in\Edg(\GR)$ 
and vertices $v_j\in\Ver(\GR)$ define the projections
\fktdefabgesetzt{\pi_\GR}{\Ab}{\Ab_\GR\ident\LG^{\elanz\Edg(\GR)}}
   {\qa}{\bigl(h_\qa(e_1),\ldots,h_\qa(e_{\elanz\Edg(\GR)})\bigr)}
to the lattice gauge theory. The topology on $\Ab$
is induced using all the $\pi_\GR$ by the topology of each 
$\LG^{\elanz\Edg(\GR)}$.

\section{A Construction Method for New Connections}
\label{sect:constr_meth}
Note that in this section we mean by "path" usually
not an equivalence class of paths, but a "genuine" path.

The main goal of this section is to provide 
a method for constructing a connection $\qa$
that only minimally, but significantly differs from a given $\qa'$.
In detail, we want to define a new connection whose parallel transport
along a given path $e$ takes a given group element $g$, but has the same parallel
transports as the older one along the other paths. However, this is obviously
impossible, because the parallel transports have
to obey the homomorphy rule. How can we find the way out? The idea goes as
follows: The only condition a connection 
has to fulfill as a map from $\Pf$ to $\LG$ 
is indeed the homomorphy property. Therefore it should be
possible to leave the parallel transports at least along those paths untouched
that do not pass any subpath of our given path $e$. Since the generalized 
connections need not fulfill any continuity condition it does not matter
"where" in $e$ the modification should be placed, e.g., whether in the 
first half or the second or perhaps in the initial point. Since we are looking
for minimal variation we try to place the modification into one single
point, say, the initial point $e(0)$. This way
all paths
that do not pass $e(0)$ can keep their parallel transports. This is
even true for
those paths that though start (or end) in the point $e(0)$, but start (or end)
in "another direction" as $e(0)$ does. Hence, we are now left
with those paths that pass an initial path of $e$. There we really have to
change the parallel transports -- we simply multiply the corresponding 
factor that changes $h_\qa(e)$ to $g$ from the left (or its inverse
from the right) to the transport of every path that starts (inversely) as $e$.
Using a certain decomposition of an arbitrary path we get the desired
construction method.

\subsection{Hyphs}
Before we state and prove the theorem 
we still need two crucial definitions and a decomposition lemma.
\bdf
Let $\gamma_1, \gamma_2\in\Pf$.

We say that $\gamma_1$ and $\gamma_2$ have the same initial segment 
(shortly: $\gamma_1 \BB \gamma_2$) iff there are non-trivial initial paths
$\gamma'_1$ and $\gamma'_2$ of $\gamma_1$ and $\gamma_2$, respectively,
that coincide up to the parametrization.

We say analogously that the final segment of $\gamma_1$ coincides
with the initial segment of $\gamma_2$ 
(shortly: $\gamma_1 \EB \gamma_2$) iff $\gamma_1^{-1} \BB \gamma_2$.
The definition of $\gamma_1 \BE \gamma_2$ and $\gamma_1 \EE \gamma_2$
should now be clear.

Iff the corresponding relations are not fulfilled, we write
$\gamma_1 \notBB \gamma_2$ etc.

\edf
\bdf
Let $\gamma$ and $\delta_i$, $i\in I$,
be a paths without self-intersections.
$\gamma$ is called \df{independent} of $D:=\{\delta_i\mid i\in I\}$
iff 
\bunum
\item
there is a $\tau\in[0,1)$ with 
$\gamma^{\tau,+}\notBB\delta_i^{\gamma(\tau),+}$ and   
$\gamma^{\tau,+}\notBE\delta_i^{\gamma(\tau),-}$ for all $i\in I$ or
\item
there is a $\tau\in(0,1]$ with 
$\gamma^{\tau,-}\notEB\delta_i^{\gamma(\tau),+}$ and   
$\gamma^{\tau,-}\notEE\delta_i^{\gamma(\tau),-}$ for all $i\in I$
\eunum
holds.\footnote{$\gamma^{\tau,+}$ is the subpath of $\gamma$ that 
corresponds to $\gamma\einschr{[\tau,1]}$; $\gamma^{\tau,-}$ 
that for $\gamma\einschr{[0,\tau]}$. Analogously,
$\delta^{x,+}$ is the subpath of $\delta$ starting in
$x$ supposed $x\in\im\delta$. (See also \cite{paper2}.) If
$\gamma(\tau)$ should not be contained in $\im\delta$ then the corresponding 
relation $\gamma^{\tau,+}\notBB\delta_i^{\gamma(\tau),+}$ etc. 
is defined to be fulfilled.} 
The point $\gamma(\tau)$ is then usually called \df{free point} of $\gamma$.

A finite set $D=\{\delta_i\}$ of paths without self-intersections
is called \df{hyph} or \df{moderately independent} iff
$\delta_i$ is independent of $D_i = \{\delta_j\mid j<i\}$.
\edf
\blem
\label{lem:fin_many_pass}
Let $\gamma\in\Pf$ and $x\in M$. Then $\gamma^{-1}(\{x\})$ is a
union of at most finitely many isolated points and 
finitely many closed intervals in $[0,1]$.
\elem
\bpf
Let $\gamma$ be (up to the parametrization) equal $\prod \gamma_i$ with
simple $\gamma'_i\in\Pf$. 
Since any $\gamma'_i$ equals (up to the parametrization) a finite
product of edges in graphs and of trivial paths, this is also true
for $\gamma$ itself.
Obviously, we can even assume w.l.o.g. that
$\gamma = \prod \gamma_i$ with $\gamma_i$ being edges in graphs or
trivial paths. (Thus, the manner of writing brackets in $\prod\gamma_i$ 
does not matter.)

The assertion of the lemma is obviously true for any $\gamma_i$ because
an edge in a graph has just been defined as non-self-intersecting
and $\gamma_i^{-1}(\{x\})$ is in the case of a trivial path either equal
$\leeremenge$ or $[0,1]$.

The case of a general $\gamma$ is now clear.
\qed
\epf
\bcorr
\label{zerleg_weg}
Let $x\in M$ be a point.
Any $\gamma\in\Pf$ can be written (up to parametrization)
as a product $\prod\gamma_i$ with $\gamma_i\in\Pf$,
such that
\bunum
\item
$\inter\gamma_i \cap \{x\} = \leeremenge$ or
\item
$\inter\gamma_i = \{x\}$.
\eunum
\ecorr
\bpf
Mark on $[0,1]$ the end points of the closed intervals and the isolated  
points of $\gamma^{-1}(\{x\})$ outside these intervals. 
We get finitely many intervals on $[0,1]$. Each one corresponds to a 
subpath $\gamma_i$ of $\gamma$. Obviously, $\prod\gamma_i$ is the desired 
decomposition of $\gamma$.
\qed
\epf
\subsection{The Construction}
How we can state the construction method.
\bconst
\label{konstr_zush}
Let $\qa\in\Ab$ and $\bweg\in\Pf$ be a path without
self-intersections.
Furthermore, let $g\in\LG$.

We now define $h:\Pf\nach\LG$.
\bunum
\item
Let $\gamma\in\Pf$ be for the moment a path that does not contain the
initial point $\bweg(0)$ of $\bweg$ as an inner point. Explicitly we have
$\inter\gamma \cap \{\bweg(0)\} = \leeremenge$.
Define

$h(\gamma) :=
\begin{cases}
  g \: h_\qa(\bweg)^{-1} \: h_\qa(\gamma) \: h_\qa(\bweg) \: g^{-1}, 
                        & \text{ for $\gamma \BB \bweg$ and $\gamma \EB \bweg$} \\
  g \: h_\qa(\bweg)^{-1} \: h_\qa(\gamma)\phantom{ \: h_\qa(\bweg) \: g^{-1}}, 
                        & \text{ for $\gamma \BB \bweg$ and $\gamma \notEB \bweg$} \\
  \phantom{g \: h_\qa(\bweg)^{-1} \: }h_\qa(\gamma) \: h_\qa(\bweg) \: g^{-1}, 
                        & \text{ for $\gamma \notBB \bweg$ and $\gamma \EB \bweg$} \\
  \phantom{g \: h_\qa(\bweg)^{-1} \: }h_\qa(\gamma)\phantom{ \: h_\qa(\bweg) 
           \: g^{-1}},  & \text{ else}        
\end{cases}$ .
\item
For every trivial path $\gamma$ set $h(\gamma) = e_\LG$.
\item
Now, let $\gamma\in\Pf$ be an arbitrary path. Decompose $\gamma$ into a 
finite product $\prod\gamma_i$ due to Corollary \ref{zerleg_weg} such that not
any $\gamma_i$ contains the point $\bweg(0)$ in the interior supposed 
$\gamma_i$ is not trivial.
Here, set $h(\gamma) := \prod h(\gamma_i)$.
\eunum
\econst
\bthm
\label{thm:constr_conn}
The map $h:\Pf\nach\LG$ from 
Construction \ref{konstr_zush} is for all $\qa$, $e$ and $g$
a homomorphism, i.e. corresponds to a connection $\qa'\in\Ab$.
\ethm
Here, $\Pf$ is the set of all {\em equivalence classes} of paths.
\bpf
\bnum{3}
\item
$h$ is a well-defined mapping from $\Pf$ to $\LG$.
\bunum
\item
Obviously, $h(\gamma') = h(\gamma'')$ if 
$\gamma'$ and $\gamma''$ coincide up to the parametrization.
Thus, we can drop the brackets in the following when we construct
multiple products of paths.
\item
Now, we show $h(\delta'\circ\delta'')=
h(\delta'\circ\delta\circ\delta^{-1}\circ\delta'')$.

Decompose 
$\delta'$, $\delta''$ and $\delta$ due to Corollary \ref{zerleg_weg}.
\bunum
\item
$\delta(0)\neq \bweg(0)$, $\delta(1)\neq\bweg(0)$ and 
$\bweg(0)\in\im\delta$

Then the decomposition of $\delta'\circ\delta''$ is equal 
$\bigl(\prod_{i=1}^{I'-1} \delta'_i\bigr) \: \gamma'''_\ast \:
 \bigl(\prod_{i=2}^{I''} \delta''_i\bigr)$ setting 
$\gamma'''_\ast := \delta'_{I'}\delta''_1$.
The decomposition of $\delta'\circ\delta\circ\delta^{-1}\circ\delta''$ is
\zgl{\bigl(\prod_{i=1}^{I'-1} \delta'_i\bigr) \: \gamma'_\ast \:
     \bigl(\prod_{i=2}^{I-1} \delta_i\bigr) \: \gamma_\ast \:
     \bigl(\prod_{i=I-1}^{2} \delta_i^{-1}\bigr) \: \gamma''_\ast \:
     \bigl(\prod_{i=2}^{I''} \delta''_i\bigr)} with
$\gamma'_\ast := \delta'_{I'} \delta_1$, 
$\gamma_\ast := \delta_{I} \delta_I^{-1}$ and
$\gamma''_\ast := \delta_1^{-1} \delta''_1$. (In the third product
the index decreases.)

A simple calculation shows that the definition above indeed yields the
same parallel transport for both paths.
\item
The other cases can be proven completely analogously.
\eunum
\item
We have as well $h(\delta'\circ\delta\circ\delta^{-1}) = 
h(\delta') =
h(\delta\circ\delta^{-1}\circ\delta')$ for all $\delta'$ and $\delta$.
\item
Since equivalent paths can be transformed into each other by a finite number
of just described transformations, we get the well-definedness.
\eunum
\item
$h$ is a homomorphism, i.e. $h$ corresponds to a  
generalized connection.

Let $\gamma$ and $\delta$ be two paths and $\prod_{i=1}^I \gamma_i$
and $\prod_{j=1}^J \delta_j$, respectively, be their decompositions as above. 
Then the decomposition of $\gamma\circ\delta$ equals
$\bigl(\prod_{i=1}^{I-1} \gamma_i\bigr) \: \gamma_\ast \:
 \bigl(\prod_{j=2}^{J} \delta_j\bigr)$
with $\gamma_\ast:=\gamma_I\delta_1$ supposed
\bunum
\item
$\gamma_I(1) \ident \delta_1(0) \neq \bweg(0)$ or 
\item
$\gamma_I(\tau)$ equals $\bweg(0)$ for all $\tau$ and so does 
$\delta_1(\tau)$.
\eunum
Otherwise the decomposition is
$\bigl(\prod_{i=1}^{I} \gamma_i\bigr)
 \bigl(\prod_{j=1}^{J} \delta_j\bigr)$ and the homomorphy is trivial
by the above definition of $h$ on general paths.

In the first case we still have to prove 
$h(\gamma_I\circ\delta_1) = h(\gamma_I) h(\delta_1)$. 
But, this can be seen quickly using the homomorphy property of $h_\qa$
and the definition above.
\qed
\enum
\epf
\brem
\bunum
\item
The theorem just proven is very well suited for the proof of the
surjectivity and the openness of $\pi_\GR:\Ab\nach\Ab_\GR$
(see below). 
In a certain sense it is 
a generalization of the proposition about the independence of loops in
\cite{paper1,a48}. This says that (for compact 
Lie groups with $\exp(\Lieg) = \LG$)
the holonomies along independent loops are even independent on the level
of regular connections. 
For instance, a set of loops is independent if each loop possesses a  
subpath called free segment that is not passed by any other loop. 
The independence proposition could be proven modifying suitably  
a given connection along those free segments, such that the
resulting holonomy becomes a certain fixed value.
In our case we do no longer need the restriction to regular connections. 
We can instead modify a 
connection "pointwise", e.g., in the point $\bweg(0)$ in
the construction above.
\item
In the compact case we will extensively use this theorem in a
subsequent paper \cite{paper4} when we prove a stratification
theorem for $\Ab$ and $\AbGb$.
\item
The theorem is valid not only for compact, but also for arbitrary 
structure groups $\LG$.
\eunum
\erem

\subsection{Consequences}
In this subsection we collect some immediate implications
given by the construction above.

First we consider the case of arbitrarily many 
paths $e_i\in E$ that are, first, independent of the corresponding
remaining paths in $E\setminus\{e_i\}$ and, second, whose end points
form a finite set containing all the free points. 
Then the parallel transports can be 
chosen freely. More precisely, we have
\bprop
\label{prop:allg_conn_constr}
Let $\qa\in\Ab$ and $I$ be a set. Let $E:=\{e_i\mid i\in I\}\teilmenge\Pf$
be a set of paths that fulfill the following conditions:
\bnum{5}
\item
\label{cond1}
$e_i$ is a path without self-intersections for all $i$.
\item
\label{cond2}
$e_i\notBB e_j$ for all $i\neq j$.
\item
\label{cond3}
$e_i\notBE e_j$ for all $i,j$.
\item
\label{cond4}
The set $V_-:=\{e_i(0)\mid i\in I\}$ of all initial points is finite.
\item
\label{cond5}
$V_-\cap\inter e_i = \leeremenge$ for all $i$.
\enum
Finally, let there be given a $g_i\in\LG$ for all $i\in I$.

Then, there exists an $\qa'\in\Ab$ such that
\bunum
\item
$h_{\qa'}(e_i) = g_i$ for all $i\in I$ and
\item
$h_{\qa'}(\gamma) = h_\qa(\gamma)$ for all $\gamma$ 
that do not 
have a subpath $\gamma'$ that fulfills $\gamma'\BB e_i$
or $\gamma'\EB e_i$ for some $i\in I$.
Especially, this holds for all $\gamma$ with
$\im\gamma\cap\bigl(\bigcup_{i\in I} \inter e_i\bigr) = \leeremenge$.
\eunum
\eprop
\bpf
First we observe that it is impossible that $\gamma\BB e_i$ and 
$\gamma\BB e_j$ for $i\neq j$, because this would imply $e_i\BB e_j$.
Analogously, $\gamma\EB e_i$ and $\gamma\EB e_j$ is impossible
for $i\neq j$.

Now we define $h:\Pf\nach\LG$ as in Construction \ref{konstr_zush}
with some modifications. Let $\gamma\in\Pf$. We decompose $\gamma$
according to the (finite number of) passages of points in $V_-$. 
Then we set for every such subpath (again denoted by $\gamma$)

$h(\gamma) :=
\begin{cases}
  g_i \: h_\qa(\bweg_i)^{-1} \: h_\qa(\gamma) 
      \: h_\qa(\bweg_j) \: g_j^{-1}, 
                     & \text{ if $\exists i:$ $\gamma \BB \bweg_i$ 
                             and $\exists j:$ $\gamma \EB \bweg_j$} \\
  g_i \: h_\qa(\bweg_i)^{-1} \: h_\qa(\gamma)
      \phantom{ \: h_\qa(\bweg_j) \: g_j^{-1}}, 
                     & \text{ if $\exists i:$ $\gamma \BB \bweg_i$ 
                             and $\forall j:$ $\gamma \notEB \bweg_j$} \\
  \phantom{g_i \: h_\qa(\bweg_i)^{-1} \: }h_\qa(\gamma) 
      \: h_\qa(\bweg_j) \: g_j^{-1}, 
                     & \text{ if $\forall i:$ $\gamma \notBB \bweg_i$ 
                             and $\exists j:$ $\gamma \EB \bweg_j$} \\
  \phantom{g_i \: h_\qa(\bweg_i)^{-1} \: }h_\qa(\gamma)
      \phantom{ \: h_\qa(\bweg_j) \: g_j^{-1}},  & \text{ else}        
\end{cases}$ 

and extend the definition by homomorphy.

As in Theorem \ref{thm:constr_conn} one easily proves that $h$ is
a well-defined homomorphism using the observation 
in the beginning of the present proof. Hence, $h=h_{\qa'}$ with some
$\qa'\in\Ab$.

Finally, one sees immediately from the definition of $h$ that
$h_{\qa'}(e_i) = g_i$ for all $i\in I$ and
$h_{\qa'}(\gamma) = h_\qa(\gamma)$ for all $\gamma$ with 
the properties above.
\qed
\epf
The preceding proposition covers both the case of webs and of graphs:
\bcorr
\label{corr:acc_gr+web}
The assumptions of Proposition \ref{prop:allg_conn_constr} are fulfilled
if $E$ is the set of all edges of a graph or the set of all curves
of a web.
\ecorr
\bpf
For finite graphs the proof is trivial. Let therefore be $E$ the set
of all curves of a web. By definition, the conditions \ref{cond1}, 
\ref{cond4} and \ref{cond5} are fulfilled as one easily checks
using the definition of a web (cf. \cite{d3}). 

To prove \ref{cond2} we assume
that $e_1\BB e_2$ for certain curves $e_1,e_2\in E$. Then we know
that $e_1(0) = e_2(0) =: p_0$, i.e., $e_1$ and $e_2$ belong to one and the same
tassel. Suppose now $\im e_1\neq\im e_2$. Then there is w.l.o.g. a $p\in M$ 
with $p\in\im e_1\setminus\im e_2$. Then, by the definition of a tassel,
in every neighbourhood of $p_0$ there is a $p'\in\im e_1\setminus\im e_2$.
But this is a contradiction to $e_1\BB e_2$. Hence, $\im e_1 = \im e_2$.
Thus, since the $e_l$ are paths without self-intersections,
there is a homeomorphism $\Pi:[0,1]\nach [0,1]$ with $e_2 = e_1 \circ \Pi$
and $\Pi(0) = 0$.
Now, due to the consistent parametrization of curves of a tassel we know
that there is a positive constant $k$ with $\Pi(\tau) = k\tau$ for all
$\tau\in[0,1]$. Because of $\Pi(1) = 1$ we get $k=1$ and $\Pi = \id$.
Thus, $e_2=e_1$. 

Finally, condition \ref{cond3} is fulfilled. In fact, let $e_i\BE e_j$. Then
we have $e_i(0) = e_j(1)$. This is obviously impossible by the definition of 
tassels and webs. 
\qed
\epf
From the proof we get immediately
\bcorr
The curves of a web form a hyph.
\ecorr
\bpf
The free point of a curve $c$ in the web is simply its initial point $c(0)$.
\qed
\epf
Now, we come to the case of arbitrary independent paths
leading to the hyphs themselves.
\bprop
\label{corr2:allg_conn_constr}
Let $\qa\in\Ab$ and $C\teilmenge\Pf$ be a set of paths without 
self-intersections. Now, let
$e\in\Pf$ be a path without self-intersections and $g\in\LG$ be
arbitrary. Furthermore, suppose that $e$ is independent of $C$.

Then there is an $\qa'\in\Ab$ such that
\bunum
\item
$h_{\qa'}(e) = g$ and
\item
$h_{\qa'}(c) = h_{\qa}(c)$ for all $c\in C$.
\eunum
\eprop
\bpf
Due to the independence of $e$ w.r.t. $C$, we have 
$e \aeqrel e^{\tau,-}\circ e^{\tau,+}$ for some 
$\tau\in[0,1]$,\footnote{If $\tau=0$
let $e^{\tau,-}$ be the trivial path and, analogously,
$e^{\tau,+}$ for $\tau=1$.} such that, w.l.o.g., $e^+:=e^{\tau,+}$ 
is a non-trivial path such that for all subpaths $c'$ of
all the $c\in C$ we have $e^+\notBB c'$ and 
$e^+\notBE c'$. Analogously to 
Proposition \ref{prop:allg_conn_constr} above
there is now an $\qa'\in\Ab$
such that with $e^-:=e^{\tau,-}$
\bunum
\item
$h_{\qa'}(e^+) = (h_{\qa}(e^-))^{-1} g$,
\item
$h_{\qa'}(c) = h_\qa(c)$ for all $c$ and
\item
$h_{\qa'}(e^-) = h_{\qa}(e^-)$.
\eunum
The last line follows, because $e$ is a path without self-intersections,
i.e., there cannot exist a subpath $e'$ of $e^-$ that 
is $\BB$ or $\EB$ to $e^+$.
Finally, we have $h_{\qa'}(e) = h_{\qa'}(e^-) h_{\qa'}(e^+) = g$.
\qed
\epf
\bcorr
\label{corr3:allg_conn_constr(hyph)}
Let $\qa\in\Ab$ be a generalized connection and 
$\hyph=\{e_1,\ldots,e_\Hyph\}\teilmenge\Pf$
be a hyph.
Furthermore, let $g_i\in\LG$, $i=1,\ldots,\Hyph$, be arbitrary.

Then there is a connection $\qa'\in\Ab$ such that
$h_{\qa'}(e_i) = g_i$ for all $i$.
\ecorr
\bpf
Use inductively the preceding corollary.
Let $\qa_0:=\qa$. Then for all $i$ choose an $\qa_i$ such that
$h_{\qa_i}(e_i) = g_i$ and $h_{\qa_i}(e_j) = h_{\qa_{i-1}}(e_j)$ for 
all $j<i$ using the assumed independence of $e_i$ w.r.t.
$\{e_j\mid j<i\}$.
Finally, set $\qa':=\qa_Y$. $\qa'$ has now the desired property.
\qed
\epf
\subsection{Surjectivity}
\bprop
\label{satz:pi_GR=surj}
$\pi_\GR:\Ab\nach\Ab_\GR$ is surjective for all graphs $\GR$.

$\pi_w:\Ab\nach\Ab_w$ is surjective for all webs $w$.

$\pi_\hyph:\Ab\nach\Ab_\hyph$ is surjective for all hyphs 
$\hyph$.\footnote{$\pi_\hyph$ is simply the map 
$\qa\auf(h_\qa(e_1),\ldots,h_\qa(e_Y))\in\LG^Y$ where $e_i$ are the
paths in $\hyph$.}
\eprop
For Lie groups with $\exp(\Lieg)=\LG$ the surjectivity of $\pi_\GR$
can also be proven
analytically showing that even $\pi_\GR\einschr\A:\A\nach\Ab_\GR$ is surjective.
In the case of webs one additionally needs compactness and semi-simplicity
of $\LG$.
But, the proof given here has the advantage that it is 
completely algebraic and needs no additional assumptions for $\LG$. 
Moreover, it uses the very constructive proposition just proven and
is valid also for hyphs.
\bpf
Let $(g_1,\ldots,g_{\elanz\Edg(\GR)})\in\LG^{\elanz\Edg(\GR)}$ be given.
Now let $\qa\in\Ab$ be the trivial connection, i.e. 
$h_\qa(\gamma) = e_\LG$ for all $\gamma\in\Pf$.
By Proposition \ref{prop:allg_conn_constr} and Corollary \ref{corr:acc_gr+web}
there is an $\qa'\in\Ab$ with $h_{\qa'}(e_i) = g_i$ for all
$i=1,\ldots,\elanz\Edg(\GR)$.

The proof in the case of webs is completely analogous, the
proof for hyphs uses Corollary \ref{corr3:allg_conn_constr(hyph)}.
\qed
\epf

\subsection{Definition of $\Ab$ Using Hyphs}
In a preceding paper \cite{paper2} we proved that in the smooth case
for a compact and semi-simple structure group $\LG$ the spaces
$\Ab_{(\infty,+)}$ and $\Ab_\Web$ of generalized connections used here
and by Baez and Sawin, respectively, are in fact homeomorphic.
Now, we will translate that proof to the case of hyphs.

First, we define a partial ordering on the set of hyphs:
$\hyph_1\leq\hyph_2$ iff every $e\in\hyph_1$ equals up to the parametrization
a finite product of paths in $\hyph_2$ and their inverses.
Then we can define $\Ab_\hyph := \Hom(\KG{\hyph},\LG)$ ($\KG{\hyph}$
being the subgroupoid of $\Pf$ generated by $\hyph$) and 
\fktdefabgesetzt{\pi_{\hyph_1}^{\hyph_2}}{\Ab_{\hyph_2}}{\Ab_{\hyph_1},}
                                         {h}{h\einschr{\KG{\hyph_1}}}
for $\hyph_1\leq\hyph_2$. We topologize $\Ab_\hyph$ identifying it with
$\LG^{\elanz\hyph}$. Obviously $\pi_{\hyph_1}^{\hyph_2}$ is always
continuous, surjective and open. So we can define
$\Ab_\HyphSet := \varprojlim_\hyph \: \Ab_\hyph$ 
as the space of generalized connections with the canonical projections
\fktdefabgesetzt{\pi_\hyph}{\Ab_\HyphSet}{\Ab_\hyph.}
                           {(h_{\hyph'})_{\hyph'}}{h_\hyph}
Using the surjectivity of $\pi_\hyph$ we get
\bprop
$\Ab_\HyphSet$ and $\Ab$ are homeomorphic in every smoothness category.
\eprop
The proof is almost literally the same as for $\Ab_\Web$ and $\Ab_{(\infty,+)}$
in \cite{paper2} and is therefore dropped here.

\section{Directedness of the Set of Hyphs}
In this section we will prove the following
\bthm
\label{thm:hyph_direct}
The set of all hyphs is directed.
\ethm
This assertion follows immediately from the more general
\bprop
\label{prop:fin_set(paths)>hyph}
Let $C\teilmenge\Pf$ be a finite set of paths without self-intersections. 
Then there is a hyph $\hyph$,
such that every $c\in C$ equals up to the parametrization 
a finite product of paths (and their inverses) in 
$\hyph$.\footnote{Consequently, for no $c\in C$ there is a 
path occuring twice in the product for $c$.}
\eprop

We will prove this theorem using induction on the number of paths
in $C$. If a path $c\in C$ would be independent
of the complement $C\setminus\{c\}$, there will be no problems. 
Therefore, we first consider the other case.
\subsection{Non-independent Paths}
In the following we often decompose paths without self-intersections
according to a finite set $P$ 
of points in the manifold $M$. This means, given some 
path $e$ we construct non-trivial
subpaths $e_i$ such that every $e_i$ starts and ends
in $P$ or $e(0)$ or $e(1)$. We obviously need only 
finitely many $e_i$ and get $e\aeqrel \prod e_i$.
\blem
\label{lem:divide_indep}
Let $e$ and $c_j$, $j\in J$, be finitely many
paths without self-intersections, such that
$e$ is not independent of $C:=\{c_j\mid j\in J\}$.

Then there are $\tau_i\in[0,1]$, $i=0,\ldots,I$, with $\tau_0=0$ and $\tau_I=1$
such that the following 
holds: After decomposing every $e$ and $c_j$ into a product of edges
$\prod_{i=0}^{I-1} e_i$ and $\prod c'_k$, respectively, according to the 
set $\{e(\tau_i)\}$, 
for every $i=0,\ldots, I-1$ 
one of the following two assertions
is true:
\bnum{2}
\item
$e_i\BB c'_k$ $\impliz$ $e_i\aeqrel c'_k$ 
and
 
$e_i\BE c'_k$ $\impliz$ $e_i\aeqrel (c'_k)^{-1}$
\item
$e_i\EB c'_k$ $\impliz$ $(e_i)^{-1}\aeqrel c'_k$ 
and 

$e_i\EE c'_k$ $\impliz$ $(e_i)^{-1}\aeqrel (c'_k)^{-1}$.
\enum
\elem
Note that here the $\aeqrel$-sign indicates that, e.g. in the first
case, $e_i$ and $c'_k$ are even equal up to the parametrization.
\bpf
\bnum{4}
\item
Let $I_{\tau,+,j}$, $\tau\in[0,1]$, contain exactly $\tau$ itself
and those $\tau'\in(\tau,1]$
for that the subpath of $e$ from $\tau$ to $\tau'$ 
is up to the parametrization equal to some subpath of $c_j$ or 
$c_j^{-1}$.
By assumption for all $\tau\in[0,1)$ there is a $j$ with
$I_{\tau,+,j}\neq\{\tau\}$.

Analogously, 
$I_{\tau,-,j}$, $\tau\in[0,1]$, contains exactly $\tau$ itself
and those $\tau'\in[0,\tau)$
for that the subpath of $e$ from $\tau'$ to $\tau$ 
is up to the parametrization equal to some subpath of $c_j$ or 
$c_j^{-1}$.
Again, by assumption for all $\tau\in(0,1]$ there is a $j$ with
$I_{\tau,-,j}\neq\{\tau\}$.

Furthermore, $I_{\tau,\pm,j}$ is everytime connected. 

Now, define 
\zgl{I_{\tau,\pm} := 
     \bigcap_{\substack{j\in J \\ I_{\tau,\pm,j}\neq\{\tau\}}}  I_{\tau,\pm,j},}
as well as $I_{0,-}:=\{0\}$ and $I_{1,+}:=\{1\}$.

What is the interpretation of such an $I_{\tau,\pm}$?
$I_{\tau,+}$, e.g., is that interval in [0,1] starting in $\tau$
such that every subpath of $c_j$ (or $c_j^{-1}$),
that starts in $e(\tau)$
as $e$ does, is even equal (up to the parametrization) 
to this subpath of $e$ at least from $e(\tau)$ to $e(\tau')$ for every
$\tau'\in I_{\tau,\pm}$.
However, note, that $I_{\tau,\pm}$ need not be a closed interval.

Observe, that $I_{\tau,\pm}$ is in each case (except for $I_{0,-}$
and $I_{1,+}$) an interval that contains
$\{\tau\}$ as a proper subset.
\item
Now, we construct a sequence $(\tau_i)$ of numbers starting with $\tau_0:=0$
as follows for all $i \geq 0$:
\bnum{4}
\item
\label{algorwegzerl1}
$\tau_{i,+}:=\sup I_{\tau_i,+}$.
\item
\label{algorwegzerl2}
$\tau_{i+1}:=\sup \{\tau\in[\tau_{i,+},1]
                    \mid I_{\tau_i,+}\cap I_{\tau,-}\neq\leeremenge\}$
\item
\label{algorwegzerl3}
$\tau_{i+1,-}$ is some number with
\bunum
\item
$\tau_{i,+}\leq\tau_{i+1,-}\leq\tau_{i+1}$,
\item
$\tau_{i+1,-}\in I_{\tau_{i+1},-}$ and
\item
$I_{\tau_i,+} \cap I_{\tau_{i+1,-},-} \neq \leeremenge$.
\eunum
\item
\label{algorwegzerl4}
$\tau_{i+\einhalb}$ is some number in $I_{\tau_i,+} \cap I_{\tau_{i+1,-},-}$.
\item
If $\tau_{i+1} = 1$ then stop the procedure.
\enum
Observe:
\bnum{4}
\item
$\tau_{i,+} > \tau_i$, because $I_{\tau_i,+}$ is a non-trivial interval.
\item
Since $I_{\tau_i,+} \cap I_{\tau_{i,+},-} \neq \leeremenge$ (by definition
of $\tau_{i,+}$), the set of all numbers $\tau$ with
$I_{\tau_i,+}\cap I_{\tau,-} \neq \leeremenge$ and $\tau\geq\tau_{i,+}$
non-empty. Consequently, it has a supremum $\tau_{i+1}\geq\tau_{i,+}$.
\item
By choice of $\tau_{i+1}$ as such a supremum there is a $\tau'\geq\tau_{i,+}$
with $\tau'\in I_{\tau_{i+1},-}$ and 
$I_{\tau_i,+}\cap I_{\tau',-}\neq\leeremenge$. Choose now 
$\tau_{i+1,-}:=\tau'$.
\item
$\tau_{i+\einhalb}$ exists obviously.
\enum
Thus, the construction above is possible.

Furthermore, we have
$\tau_i \leq \tau_{i+\einhalb} \leq \tau_{i,+} 
        \leq \tau_{i+1,-} \leq \tau_{i+1}$ and $\tau_i < \tau_{i+1}$.
\item
Now, assume that there is no $N\in\N$ with $\tau_N=1$.
Then $(\tau_i)_{i\in\N}$ is a strictly increasing sequence with
values in $[0,1)$, i.e. $\tau_i\gegen\tau\in(0,1]$ for $i\gegen\infty$,
and we have $\tau_i < \tau$ for all
$i\in\N$.

Let $\tau'\in I_{\tau,-}$ with $\tau'<\tau$. Then there is an $n\in\N$
with $\tau'\leq\tau_n<\tau$. Now we have
$I_{\tau_n,+}\cap I_{\tau,-}\neq\leeremenge$, because, e.g.,
$\tau_n$ is contained in this set.
But, from this we get together the step 
\ref{algorwegzerl2} above, that $\tau\leq\tau_{n+1}$.
This is a contradiction to $\tau>\tau_{n+1}$.

Consequently, there is an $N\in\N$ with $\tau_N=1$.
\item
Now, the desired parameter values are
$\tau_i$, $\tau_{i+\einhalb}$ and 
$\tau_{i+1,-}$ for $i=0,\ldots,N-1$ as well as $\tau_N$. 
Divide the edges $e$ and $c_j$ according to the 
set of all those $e(\tau_{\ldots})$.
We have (if two subsequent vertices $e(\tau_{\ldots})$
are equal, we drop the correspondent (trivial) subpaths
$e_{\ldots}$ and $c'_{\ldots}$):
\bnum{3}
\item
$e_i\BB c'_k$ $\impliz$ $e_i\aeqrel c'_k$ 
and
 
$e_i\BE c'_k$ $\impliz$ $e_i\aeqrel (c'_k)^{-1}$;
\item
$e_{i+\einhalb}\EB c'_k$ $\impliz$ $(e_{i+\einhalb})^{-1}\aeqrel c'_k$ 
and 

$e_{i+\einhalb}\EE c'_k$ $\impliz$ $(e_{i+\einhalb})^{-1}\aeqrel (c'_k)^{-1}$;
\item
$e_{i+1,-}\EB c'_k$ $\impliz$ $(e_{i+1,-})^{-1}\aeqrel c'_k$ 
and 

$e_{i+1,-}\EE c'_k$ $\impliz$ $(e_{i+1,-})^{-1}\aeqrel (c'_k)^{-1}$.
\enum
We only show the first item, the two other ones can be proven analogously.

Let $e_i\BB c'_k$. Since $c'_k$ is a subpath of a $c_j$,
we have $I_{\tau_i,+,j}\neq\{\tau_i\}$. From
$I_{\tau_i,+,j}\obermenge I_{\tau_i,+}\obermenge[\tau_i,\tau_{i+\einhalb}]$
we get now $e_i$ equals (up to the parametrization) a subpath of 
$c_j$ starting in $e(\tau_i)$. But, since $c_j$ has no self-intersections
and is divided according to $e(\tau_i)$ and $e(\tau_{i+\einhalb})$ (and
other vertices that are not contained in $\im e_i$),
we have $e_i$ even equals $c'_k$ up to the parametrization.

In the case $e_i\BE c'_k$ we conclude analogously
using $e_i\BB (c'_k)^{-1}$.
\qed
\enum
\epf
\subsection{Proof of Proposition \ref{prop:fin_set(paths)>hyph}}
\bpf[Proposition \ref{prop:fin_set(paths)>hyph}]
\bunum
\item
First of all we decompose all $c_i$ according to the set
$V:=\{c_i(0)\}_i\cup\{c_i(1)\}_i$ of all end points. Thus, we
get a finite set $C'$ of paths without self-intersections, whereas
every $c\in C$ equals up to the parametrization a finite product 
of paths $c'\in C'$ and their inverses and where no end point
of a path $c'$ is contained in the interior of another path in $C'$.

Consequently, we can w.l.o.g. assume that our set $C$ in the 
proposition is of that type.
\item
Now, we consider $c_1\in C$. 

\bnum{2}
\item
In the case that $c_1$ is already independent
of $\{c_j\mid j>1\}$ we need not decompose $c_1$; we simply set 
$c_{i,1}:=c_i$ and $I_i:=1$ for all $i$.
\item
In the other case we use Lemma
\ref{lem:divide_indep} and get certain
paths $e_k$ (w.l.o.g. such that 
$c_1\aeqrel e_1\circ\cdots\circ e_{I_1}$) such that 
every $c_j$ is a product of the $e_k$ (and their inverses) and such that
the $e_k$, $k\in[1,I_1]$, are independent of the remaining paths.
Now, we set $c_{1,k}:=e_k$ for all $k\in[1,I_1]$.
Analogously, we define $c_{i,l}$ for $i>1$ being that $e_k$
that (or whose inverse) is used at the $l$th position in the product
for $c_i$, after we cancelled all $e_k$ occuring in $c_1$, and
denote the number of factors left by $I_i$.\footnote{Example:
$c_1 = e_1 e_2 e_3$, $c_2 = e_1^{-1} e_4 e_3 e_5^{-1}$ and $c_3 = e_2^{-1}$.
Then we have
$I_1 = 3$, $I_2=2$, $I_3=0$ and
$c_{1,1} = e_1$, $c_{1,2} = e_2$, $c_{1,3} = e_3$, 
$c_{2,1} = e_4$ and $c_{2,2} = e_5$.} 
\enum
Per constructionem, $c_{1,l}$ is independent of 
$\{c_{i,l'}\mid i>1 \text{ or } l\neq l'\}$.
Note, moreover, that the set of end points of the $c_{i,l}$ is
again disjoint to the interiors of these paths.
Finally, we set $C_1:=\{c_{i,l}\mid i>1\}$.
\item
Now, we decompose the paths $c_{2,l}\in C_1$ (if $I_2\neq 0$).

We start with $c_{2,1}$. If it is not independent of the 
$\{c_{i,l}\in C_1\mid i>2\text{ or }l\neq 1\}$, then 
decompose it again by Lemma \ref{lem:divide_indep} by certain
independent paths $e'_k$.
We get as before 
$c_{2,1} \aeqrel c_{2,1,1}\circ\cdots\circ c_{2,1,I_{2,1}}$
and a certain set $C_{2,1}$ that collects all paths
used for the decomposition of $c_{i,l}$ with $i>2$.
But, note that $c_{2,l}$ is {\em not} decomposed for $l\neq 1$ by that
procedure.

Afterwards, we decompose $c_{2,2}$ (w.r.t. $C_{2,1}$) and so on.

Summa summarum, we get paths $c_{2,l,m_l}$ with
$c_{2,l} \aeqrel \prod_{m_l} c_{2,l,m_l}$ and a set $C_2:=C_{2,I_2}$
collecting all the paths that $c_{i,l}$ with $i>2$ is decomposed 
into, but that are not used in the decomposition of $c_{2,l}$.
By the construction,
$c_{2,l,m_l}$ is independent of
$\{c_{2,l',m'_{l'}}\mid l\neq l' \text{ or } m_l\neq m'_{l'}\}\cup C_2$.
\item
In the next step, we first collect all paths in $C_2$
that are used for the decomposition of $c_3$. After renumbering
these paths by $c_{3,1},\ldots,c_{3,I_3}$ we can again apply
the previous step.
\item
Inductively, we get an ordered set
\zgl{C^\ast = \{c_{N,1,1}, \ldots, c_{N,I_N,M_{N,I_N}}; \ldots \ldots;
                c_{2,1,1}, \ldots, c_{2,I_2,M_{2,I_2}}; 
                c_{1,1}, \ldots, c_{1,I_1}\}}
of paths that is by construction moderately independent,
consequently a hyph, and that admits a factorization of every $c_i\in C$
into a product of paths in $C^\ast$ of the desired type.
\qed
\eunum
\epf
\subsection{Open Problem}
In contrast to the case of graphs or webs we need for the definition
of the independence in the case of hyphs an ordering among
the paths collected in a hyph.
Thus, it would be -- at least for technical reasons -- desirable to 
solve the following open problem: 
Does there exist for every given finite set $C$ of paths a set $E$
of strongly independent paths, such that every path in $C$ is a product
of paths in $E$ and their inverses? Strongly independent means here
that every path in $C$ is independent of the remaining paths in $C$.
We indicate the problems that arised when we tried to prove the following
answers:
\bvl
\iitem["Yes":]
The induction used for the proof of Proposition 
\ref{prop:fin_set(paths)>hyph} cannot be reused. The problem is the following.
Suppose we have decomposed the first path $c_1$ in $C$ w.r.t. to the remaining
paths as above. Then we decompose (the subpaths of) the second path $c_2$ in $C$
w.r.t. the others. Now, it is possible that vertices used in this procedure
for the division of $c_2$
lie on $c_1$ again. Thus, $c_1$ would now be divided once more -- with the
effect that sometimes subpaths of $c_1$ are created that do not fulfill the
independence condition. (Remember that independence means existence of {\em 
one} point in a path with the independence-of-germs condition above.) 
Hence, we have to divide the respective path again. But, now we could end up
in a never-ending procedure that creates an infinite number of subpaths. 
\iitem["No":]
It would be enough to present one counterexample. 
But, up to now, none of the examples we checked lead to a 
contradiction.
\evl
\section{Openness of $\pi_\GR$}
\label{app:surj+off(pi_GR)}
\bprop
$\pi_\GR:\Ab\nach\Ab_\GR$ is open for all graphs $\GR$.
\eprop
\bpf
We have to show: $\pi_\GR(V)$ is open for all elements $V$ of a basis
of $\Ab$, i.e., 
$\pi_\GR(\pi^{-1}_{\GR'_1}(W_1)\cap\ldots\cap\pi^{-1}_{\GR'_I}(W_I))$ 
is open for all graphs $\GR'_i$ and all elements $W_i$ of a 
basis of $\Ab_{\GR'_i} = \LG^{\elanz\Edg(\GR'_i)}$.
But, a basis hereof is given by all sets of the type
$W_{i,1}\kreuz\cdots\kreuz W_{i,\elanz\Edg(\GR'_i)}$ with
open $W_{i,n_i}\teilmenge\LG$. Now we have
\zgl{\pi_\GR(\pi^{-1}_{\GR'_1}(W_1)\cap\ldots\cap\pi^{-1}_{\GR'_I}(W_I)) =
     \pi_\GR\bigl(\bigcap_{i=1}^I\bigcap_{j_i=1}^{\elanz\Edg(\GR'_i)}
                 \pi^{-1}_{e_{i,j_i}}(W_{i,j_i})\bigr).}
(W.l.o.g. we assumed that none of the $\GR'_i$ consists of a single
vertex.)

Let us therefore prove the openness of all sets
of the type 
\zgl{\pi_\GR\bigl(\bigcap_{j=1}^J \pi^{-1}_{c_j}(W_j)\bigr)}
with edges $c_j$ and open $W_j\teilmenge\LG$.

Let us denote the edges of $\GR$ by $e_i$ and set $E:=\{e_i\}$ and
$C:=\{c_j\}$.
\bnum{4}
\item
Suppose first that there is an $e\in E$ that is independent 
of $C$. Then it is obviously independent of $C\cup(\Edg(\GR)\setminus\{e\})$.
We will show that 
\zgl{\pi_\GR\bigl(\bigcap_{j=1}^J \pi^{-1}_{c_j}(W_j)\bigr) = 
     \pi_{\GR\setminus\{e\}}
          \bigl(\bigcap_{j=1}^J \pi^{-1}_{c_j}(W_j)\bigr)\kreuz\LG.}
\begin{SubSet}
Trivial.
\end{SubSet}
\begin{SuperSet}
Let $(\vect g,g)\in 
     \pi_{\GR\setminus\{e\}}\bigl(\bigcap_{j=1}^J \pi^{-1}_{c_j}(W_j)\bigr) 
     \kreuz \LG$.

Hence, there is an $\qa\in\bigcap_{j=1}^J \pi^{-1}_{c_j}(W_j)$
with $\pi_{\GR\setminus\{e\}}(\qa) = \vect g$.

Due to Proposition \ref{corr2:allg_conn_constr} there is an $\qa'\in\Ab$
fulfilling
\bunum
\item
$h_{\qa'}(e_i) = h_\qa(e_i)$ for all $e_i\neq e$, i.e.
$\vect g = \pi_{\GR\setminus\{e\}}(\qa) = \pi_{\GR\setminus\{e\}}(\qa')$,
\item
$h_{\qa'}(c_j) = h_\qa(c_j)$ for all $j = 1,\ldots, J$,
i.e. $\qa'\in\pi^{-1}_{c_j}(W_j)$ for all $j$, and
\item
$h_{\qa'}(e) = g$.
\eunum
With this we have 
$\pi_\GR(\qa') = \bigl(\pi_{\GR\setminus\{e\}}(\qa'),\pi_e(\qa')\bigr)
               = (\vect g, g)$,
i.e.
\zgl{(\vect g, g) \in 
      \pi_\GR\bigl(\bigcap_{j=1}^J \pi^{-1}_{c_j}(W_j)\bigr).}
\end{SuperSet}
\item
Successively applying the preceding step we get
\zgl{\pi_\GR\bigl(\bigcap_{j=1}^J \pi^{-1}_{c_j}(W_j)\bigr) = 
     \pi_{\GR_0}\bigl(\bigcap_{j=1}^J \pi^{-1}_{c_j}(W_j)\bigr) \kreuz \LG^n.}
Here $n$ denotes the number of edges $e$ of $\GR$ that are independent
of $C$.
$\GR_0$ denotes that graph that arises
from $\GR$ by removing all such edges. 
\item
Since every edge $e$ in $\GR_0$ is not independent of $C$, we can divide
$e_1$ and the $c_j\in C$ as in Lemma \ref{lem:divide_indep} and get
paths $e_{1,1},\ldots,e_{1,n_1}$ and $c_{j,1},\ldots,c_{j,m_j}$.
We collect the $c_{\ldots}$ into $C_1\teilmenge\Pf$. Since the $e_i$
are edges of one and the same graph, $e_i$ (for $i>1$) is still
not independent of $C_1$. We again use Lemma \ref{lem:divide_indep},
now for decomposing $e_2$ and the paths in $C_1$. We get paths
$e_{2,1},\ldots,e_{2,n_2}$ and a $C_2\teilmenge\Pf$. Successively, we
decompose all $e_i$ and $C_{i-1}$ getting $e_{k,i_k}$ and 
$c'_l\in C'\teilmenge\Pf$,
such that for every $i$ and $k_i$ 
one of the following two assertions
is true:
\bnum{2}
\item
$e_{i,k_i}\BB c'_l$ $\impliz$ $e_{i,k_i}\aeqrel c'_l$ 
and
 
$e_{i,k_i}\BE c'_l$ $\impliz$ $e_{i,k_i}\aeqrel (c'_l)^{-1}$
\item
$e_{i,k_i}\EB c'_l$ $\impliz$ $(e_{i,k_i})^{-1}\aeqrel c'_l$ 
and 

$e_{i,k_i}\EE c'_l$ $\impliz$ $(e_{i,k_i})^{-1}\aeqrel (c'_l)^{-1}$.
\enum
To reduce the technical efforts we first invert all $e_{i,k_i}$ that
fulfill the second assertion.
Afterwards, we invert $c'_l$ if
it is equivalent to an $(e_{i,k_i})^{-1}$. This is possible, because 
there is at most one such edge $e_{\ldots}$.

It is clear, that the $e_{i,k_i}$ span a graph $\GR'\geq\GR_0$,
and we know from the construction that no $\inter c'_l$
contains a vertex of $\GR'$. Furthermore, every $c_j$ is equivalent to
a finite product of $c'_l$ (or its inverse). 
The factors used for $c_j$ (again denoted by $c_{j,l_j}$) span a graph
$\GR_j$, as well.
Thus, we have 
$\pi_{\GR_0} = \pi_{\GR_0}^{\GR'} \pi_{\GR'}$
and $\pi^{-1}_{c_j} = \pi^{-1}_{\GR_j} (\pi_{c_j}^{\GR_j})^{-1}$.

Finally, $(\pi_{c_j}^{\GR_j})^{-1}(W_j)$ is open in $\LG^{m_j}$ by continuity,
i.e., a union of sets of the type $W_{j,1}\kreuz\cdots\kreuz W_{j,m_j}$.
Thus, $\pi_{\GR_0}\bigl(\bigcap_{j=1}^J \pi^{-1}_{c_j}(W_j)\bigr)$
is the union of sets of the type
$\pi_{\GR_0}^{\GR'} \pi_{\GR'}
 \bigl(\bigcap_{j=1}^J \bigcap_{l_j=1}^{m_j}
       \pi^{-1}_{c_{j,l_j}}(W_{j,l_j})\bigr)$.
\item
Due to the openness of $\pi_{\GR_0}^{\GR'}$ (see \cite{paper2}) 
it is sufficient to prove the openness of 
$\pi_{\GR'}\bigl(\bigcap_{l=1}^L \pi^{-1}_{c_l}(W_l)\bigr)$ whenever
the following holds:
\bnum{2}
\item
$\GR'$ is a graph and $C' = \{c_l\}$ is a finite set of
paths without self-intersections,
\item
$\inter c_l \cap \Ver(\GR') = \leeremenge$,
\item
($e\BB c_l$ $\impliz$ $e\aeqrel c_l$) 
and $e\notBE c_l$ for all $l$ and 
for every edge $e$ of the graph $\GR'$ and 
\item
$W_l\teilmenge\LG$ is open for all $l$.
\enum
We will prove for non-empty left hand side
\znumgl{\pi_{\GR'}\bigl(\bigcap_{l=1}^L \pi^{-1}_{c_l}(W_l)\bigr) 
      = \dirprod_{e_k\in\Edg(\GR')}\bigl(\bigcap_{c_l\in C(e_k)} W_l\bigr),
        \label{openequal}}
where $C(e_k)\teilmenge C'$ contains exactly those $c_l\in C'$
that are (up to the parametrization) equal to $e_k$ or $e_k^{-1}$.
Since the right hand side is obviously open, the openness is proven
if \eqref{openequal} is.

\begin{SubSet}
Let $\vec g\in\pi_{\GR'}\bigl(\bigcap_{l=1}^L \pi^{-1}_{c_l}(W_l)\bigr)$, i.e.,
there is an $\qa\in\Ab$ with $\pi_{e_k}(\qa) = g_k$ for all $k$ and
$\pi_{c_l}(\qa)\in W_l$ for all $c_l\in C'$.
From this follows $g_k\in W_l$ for all $c_l\in C(e_k)$ and so 
$\vec g\in\dirprod_{e_k\in\Edg(\GR')}\bigl(\bigcap_{c_l\in C(e_k)} W_l\bigr)$.
\end{SubSet}
\begin{SuperSet}
Let $\vec g\in\dirprod_{e_k\in\Edg(\GR')}\bigl(\bigcap_{c_l\in C(e_k)} W_l\bigr)$.
Choose an $\qa_0\in\Ab$ with $\pi_{c_l}(\qa_0) \in W_l$ for all $c_l$.
By assumption
every $e_k$
is independent of $C'\setminus (\bigcup_k C(e_k))$ and so 
by Proposition \ref{corr2:allg_conn_constr}
there exists an $\qa\in\Ab$
such that 
\bunum
\item
$\pi_{e_k}(\qa) = g_k$ for all $k$ and
\item
$\pi_{c_l}(\qa) = \pi_{c_l}(\qa_0)$ for all $c_l$ that are not
equal (up to the parametrization) an $e_k$.
\eunum
Thus, we have $\pi_{c_l}(\qa) \in W_l$
for all $c_l\in C(e_k)$. Consequently,
$\vec g\in\pi_{\GR'}\bigl(\bigcap_{l=1}^L \pi^{-1}_{c_l}(W_l)\bigr)$.
\qed
\end{SuperSet}
\enum
\epf

\section{Induced Haar Measure}
\label{sect:AL-measure}
In this section we will show that thanks to the directedness of the set
of hyphs an induced Haar measure can be defined for arbitrary smoothness
assumption for the paths. Our definition covers that of Ashtekar and
Lewandowski for graphs in the analytic category \cite{a48} as well as that
of Baez and Sawin for webs in the smooth category \cite{d3}.

Throughout this section, $\LG$ is a {\em compact} Lie group.
\subsection{Cylindrical Functions}
In this subsection we will investigate the algebra of continuous
functions on $\Ab$. Particulary nice is the dense subalgebra of 
the so-called cylindrical functions \cite{a48,a30}. These are functions 
depending only on the parallel transports along a finite number of paths.
\bdf
A function $f\in C(\Ab)$ is called \df{genuine cylindrical function}
on $\Ab$ iff there is a graph $\GR$ and a continuous function
$f_\GR\in C(\Ab_\GR)$ with $f = f_\GR\circ\pi_\GR$.
The set of all genuine cylindrical functions is denoted by $\Cyl_0(\Ab)$.
\edf
Obviously, $\Cyl_0(\Ab)$ is $\ast$-invariant.
But, since for two finite graphs there need not exist a third one
containing both, the sum as well as the product of two cylindrical functions
is no longer a cylindrical function in general.
Therefore we enlarge the definition above to hyphs.
\bdf
A function $f\in C(\Ab)$ is called \df{cylindrical function}
on $\Ab$ iff there is a hyph $\hyph$ and a continuous function
$f_\hyph\in C(\Ab_\hyph)$ with $f = f_\hyph\circ\pi_\hyph$.
The set of all cylindrical functions is denoted by $\Cyl(\Ab)$.
\edf
\blem
$\Cyl(\Ab)$ is a normed $\ast$-algebra containing $\Cyl_0(\Ab)$.
\elem
\bpf
$\Cyl(\Ab)$ is obviously closed w.r.t. scalar multiplication
and involution. It remains to prove that it is closed w.r.t. to addition
and multiplication.

Let $f' = f'_{\hyph'}\circ\pi_{\hyph'}$ and $f'' = f''_{\hyph''}\circ\pi_{\hyph''}$.
By Theorem \ref{thm:hyph_direct} there is a 
hyph $\hyph$ with $\hyph\geq\hyph',\hyph''$.
Thus we have
$f' + f'' = f'_{\hyph'}\circ\pi_{\hyph'}^\hyph\circ\pi_\hyph +
            f''_{\hyph''}\circ\pi_{\hyph''}^\hyph\circ\pi_\hyph
          = (f'_{\hyph'}\circ\pi_{\hyph'}^\hyph +
             f''_{\hyph''}\circ\pi_{\hyph''}^\hyph)\circ\pi_\hyph\in\Cyl(\Ab)$.
Analogously, $f'\cdot f''\in\Cyl(\Ab)$.
\qed
\epf
\bprop
\label{Satz:Cyl(Ab)_dicht_C(AB)}
$\Cyl(\Ab)$ is dense in $C(\Ab)$.
\eprop
\bpf
The assertion follows from the Stone-Weierstra\ss\ theorem:
\bunum
\item
$1\in\Cyl(\Ab)$, whereas $1:\Ab\nach\C$ is the function $1(\qa) := 1$.
\item
$\Cyl(\Ab)$ seperates the points of $\Ab$:\footnote{We prove even 
$\Cyl_0(\Ab)$ seperates the points of $\Ab$.}

Let $\qa_1,\qa_2\in\Ab$ with $\qa_1\neq\qa_2$.
Thus, there is a graph $\GR$ with $\pi_\GR(\qa_1) \neq \pi_\GR(\qa_2)$.

Since $\Ab_\GR\ident\LG^{\elanz\Edg(\GR)}$ 
is a manifold, hence completely regular, the continuous
functions on $\Ab_\GR$ separate the points of $\Ab_\GR$ \cite{Kelley}.
This means there is an $f_\GR\in C(\Ab_\GR)$
with $f_\GR(\pi_\GR(\qa_1))\neq f_\GR(\pi_\GR(\qa_2))$.

Due to $f_\GR\circ\pi_\GR\in\Cyl(\Ab)$, 
$\Cyl(\Ab)$ separates the points of $\Ab$.
\qed
\eunum
\epf
\subsection{The Induced Haar Measure on $\Ab$}
According to the Riesz-Markow theorem measures on a compact
Hausdorff space are in one-to-one correspondence to linear,
continuous, positive functionals on the function algebra over that space.
We get
\bprop
For every linear, continuous, positive functional $F$ on $C(\Ab)$
there is a unique regular Borel measure $\mu$ on $\Ab$, such that
\fktdefabgesetzt{F}{C(\Ab)}{\C.}{f}{\int_\Ab f\: d\mu}
\eprop
Due to the denseness of $\Cyl(\Ab)$ in $C(\Ab)$ it is sufficient to define
an appropriate functional on $\Cyl(\Ab)$ and to extend this continuously
to a functional on $C(\Ab)$. One possibility is 
to replace the integration of functions $f_\hyph\circ\pi_\hyph$ over $\Ab$ by
the integration of $f_\hyph$ over $\Ab_\hyph = \LG^{\elanz\hyph}$. But,
on $\LG^{\elanz\hyph}$ there is a "canonical" measure,
the Haar measure. Hence, we define (cf. \cite{a48}):
\bdf
Let $f\in\Cyl(\Ab)$. Define
$F_0(f) := \int_{\Ab_\hyph} f_\hyph\: d\mu_\Haar$, 
if $f_\hyph\circ\pi_\hyph = f$, and 
extend $F_0$ continuously to a functional $F$ on $C(\Ab)$.
\edf
\bprop
$F:C(\Ab)\nach\C$ is a well-defined, linear, continuous, positive
functional on $C(\Ab)$.

Furthermore, there is a unique Borel measure $\mu_0$ on $\Ab$ with
$F(f) = \int_\Ab f\:d\mu_0$ for all $f\in C(\Ab)$.
\eprop
\bdf
The measure $\mu_0$ of the preceding proposition is called \df{induced
Haar measure} or \df{Ashtekar-Lewandowski measure} on $\Ab$.
\edf
\bpf
\bunum
\item
$F_0$ ist well-defined.

Let $f$ be cylindrical w.r.t. $\hyph'$ and $\hyph''$. Then $f$ is
again cylindrical w.r.t. $\hyph$, if $\hyph$ is some hyph containing
$\hyph'$ and $\hyph''$.
The existence of such an $\hyph$ is guaranteed by Theorem
\ref{thm:hyph_direct}.
Hence, it is sufficient to prove 
$\int_{\Ab_\hyph} f_\hyph\: d\mu_\Haar 
     = \int_{\Ab_{\hyph'}} f_{\hyph'}\: d\mu_\Haar$
for all $\hyph\geq\hyph'$.

Let now $\hyph\geq\hyph'$. Then every path $e'_i$ of $\hyph'$ can be written
as a product $\prod_{k_i} {e_{j(k_i,i)}^{\pm 1}}$
of paths in $\hyph$ (and their inverses). By the moderate independence
of hyphs there is a path $e_{K(i)}$ for every $i$, such that
$e_{K(i)}$ occurs exactly once in the decomposition of $e'_i$ and
does not occur in that of $e'_{i'}$ with $i'<i$.
Now we have 
($n:=\elanz \hyph$ and $n':=\elanz\hyph'$)
\bgl[2ex]
\zurueck 
 &   & \int_{\Ab_{\hyph}} f_{\hyph}\: d\mu_\Haar \s
\zurueck 
 & = & \int_{\LG^n} f_{\hyph}(g_1,\ldots,g_{n}) \: d\mu_\Haar \s
\zurueck 
 & = & \int_{\LG^n}  
       f_{\hyph'}\Bigl(\prod_{k_1} {g_{j(k_1,1)}^{\pm 1}},\ldots,
                       \prod_{k_{n'}} {g_{j(k_{n'},n')}^{\pm 1}}\Bigr)
       \: \prod d\mu_\Haar  \\
\zurueck 
 &   & \erl{$f_\hyph = f_{\hyph'}\circ\pi_{\hyph'}^{\hyph}$ 
            and decomposition of $e'_i$} \s
\zurueck 
 & = & \int_{\LG}\cdots\int_{\LG}  
       f_{\hyph'}(\cdots g_{K(1)}^{\pm 1}\cdots,\ldots,
                  \cdots g_{K(n')}^{\pm 1}\cdots) 
       \: d\mu_{\Haar,1}\cdots d\mu_{\Haar,n} \\
\zurueck 
 &   & \erllang[0.65\textwidth]%
               {The dots in $\cdots g_{K(l)}^{\pm 1}\cdots$ denote always
                a product of $g_j^{\pm 1}$ with $j\neq K(l')$ for all $l'>l$.}
       \s
\zurueck 
 & = & \int_{\LG}\cdots\int_{\LG}  
       f_{\hyph'}(g_1,\ldots,g_{n'}) 
       \: d\mu_{\Haar,1}\cdots d\mu_{\Haar,n'} \\
\zurueck 
 &   & \hspace*{-2ex}\erl{Translation and inversion invariance, 
            normalization of the Haar measure} 
       \s
\zurueck 
 & = & \int_{\Ab_{\hyph'}} f_{\hyph'}\: d\mu_\Haar. 
\egl
\item
$F_0$ is continuous due to $\betrag{F_0(f)}\leq\norm{f_\hyph}=\norm{f}$.
The last equality follows from the surjectivity of $\pi_\hyph$, see
Proposition \ref{satz:pi_GR=surj}.
\item
$F_0$ is obviously linear and positive.
\item
Hence, $F$ is a well-defined, linear, continuous, positive functional 
on $C(\Ab)$. 
\item
Due to the Riesz-Markow theorem there is a unique Borel measure
$\mu_0$ on $\Ab$ with $F(f) = \int_\Ab f \: d\mu_0$.
\item
$F$ is strictly positive.

Let $f\in C(\Ab)$, $f\neq 0$, and $k:=f^\ast f\in C(\Ab)$. Then 
$U := k^{-1}((\einhalb\norm{k},\infty))$ is open and non-empty. Thus,
there is a hyph $\hyph$ and an open, non-empty $U_\hyph$ with
$\pi_\hyph^{-1}(U_\hyph) \teilmenge U$. Since every open 
non-empty subset of a compact Lie group has non-vanishing Haar 
measure,\footnote{Let $U\teilmenge\LG$ be open, non-empty. Then 
$\{Ug\mid g\in\LG\}$ is a covering of $\LG$. Since $\LG$ is compact,
there are only finitely many $g_i$, such that $\bigcup_{i=1}^n Ug_i = \LG$.
Due to the translation invariance of the Haar measure we have
$\mu(U) = \inv n \sum\mu(U g_i) \geq \inv n \mu(\LG) > 0$.}
we have
\bgl[2ex]
F(f^\ast f) & = & \int_\Ab k \: d\mu_0
                  \breitrel\geq \int_U \einhalb\norm{k} \: d\mu_0 \s
         & \geq & \einhalb\norm{k} \int_{\pi_\hyph^{-1}(U_\hyph)} 1 \:d\mu_0
                  \breitrel= \einhalb\norm{k} \int_{U_\hyph} 1 \:d\mu_\Haar\s
            & = & \einhalb\norm{k} \mu_\Haar(U_\hyph) \breitrel> 0.
\keinseitenumbr
\egl
\keinseitenumbr
\qed
\eunum
\epf

\section{Discussion}
In this paper we investigated for some examples how the theory of 
generalized connections depends on the chosen smoothness category for
the paths used in the construction of $\Ab$. The most important theorem yields
that in every case an induced Haar measure can be defined. 
But, there are some problems that depend very crucially on the smoothness
of the paths. So let us resume the discussion of the beginning of this paper:
What could be a good choice of smoothness conditions?

One decisive point is the denseness of the classical (smooth) connections
in the space $\Ab_{(r)}$. In the case of compact structure groups $\LG$
the denseness has been proven for the 
immersive smooth \cite{d3,e46} and piecewise analytic category \cite{e8}.
However, in the first case \cite{d3} the space $\Ab_\Web$ was defined not
by $\varprojlim_w \Ab_w$, but by $\varprojlim_w \A_w$ where $\A_w$ 
(being a Lie subgroup of $\LG^{\elanz w}$)
denotes the image of the space $\A$ of {\em regular} connections under the 
map $\pi_w \ident h_{c_1} \kreuz \cdots \kreuz h_{c_W}$. Thus, the denseness
follows immediately by the directedness of the set of webs (cf. Appendix
\ref{app:lem_dense_projlim}).
Supposed, $\LG$ is in addition semi-simple, Lewandowski and Thiemann
\cite{e46} proved that $\A_w = \Ab_w = \LG^{\elanz w}$ which implies
that $\A$ is also dense in our $\Ab_{(\infty,+)}$. Up to now, we do not
know whether this is true for arbitrary Lie groups. However, $\A$ is 
definitely {\em not} dense in the space $\Ab_{(r)}$ for non-immersed paths. 
Let, e.g., $\gamma$ be an immersed path without self-intersections 
and $\gamma'(\tau):=\gamma(\tau^2)$. Then
$\gamma'$ is not equivalent to $\gamma$ (cf. \cite{paper2}) and
not an immersion. But, obviously $h_\gamma(A) = h_{\gamma'}(A)$ for all $A\in\A$.
Consider now two elements $g,g'\in\LG$ and corresponding 
disjoint open neighbourhoods $U, U'\teilmenge\LG$. We see that
$\hyph:=\{\gamma,\gamma'\}$ is a hyph and so
$\pi_{\gamma}^{-1}(U) \cap \pi_{\gamma'}^{-1}(U') = \pi_{\hyph}^{-1}(U\kreuz U')$
is non-empty and open, but contains no regular $A$. So $\A$ is not dense
in $\Ab_{(r)}$. 

Since this is, in fact, very unsatisfactory, we should look for other
possibilities for the definition of the set $\Pf$ for non-immersive paths.
The probably easiest way should be to redefine the equivalence relation 
between paths. Why should non-self-intersecting paths $\gamma$ and
$\gamma'$
only be equivalent if they coincide up to a piecewise $C^r$-transformation?
Perhaps we should use a definition of the following kind: 
$\gamma\aeqrel\gamma'$ iff $h_A(\gamma) = h_A(\gamma')$ for all $A\in\A$ --
maybe at least provided $\im\gamma = \im\gamma'$. This one is quite similar
to that used originally in \cite{a72,a48}. 
On the one hand, we expect that all the constructions made in this paper and
its predecessor \cite{paper2} will still go through. But, on the other hand,
even for that definition we do not see that it saves the desired density
property in more cases than described above.

What other questions discussed in the Ashtekar framework could be
touched by the choice of $\Pf$? One area we mentioned above -- the diffeomorphism
invariance of quantum gravity. Here, obviously, we have to admit at least smooth
paths. Another problem is quantum geometry. For instance, the definition
of the area operator \cite{a13} enforced the usage of at most 
the analytic category.
There one has to calculate sums over intersection points
of spin networks with surfaces. But, since there can exist
infinitely many such points when working with smooth paths, 
these sums can be infinite. 
This problem could be solved
if there would exist 
for every fixed surface $S$
in $M$ a basis of $L_2(\Ab,\mu_0)$, 
such that every base element has only finitely many intersection 
points with $S$. But this seems very unlikely.

\section{Acknowledgements}
The author was supported by the Max-Planck-Institut f\"ur Mathematik
in den Naturwissenschaften in Leipzig.

\anhangengl
\section{Additional Results for $\AbGb$}
In this appendix we give three corollaries about assertions
that can be proven not only for $\Ab$, but also for $\AbGb$.
For the definition of $\AbGb$ and the used notation we refer
to \cite{paper2}.
\bcorr
\label{folg:pi_GR(agb)=surj}
$\pi_\GR:\agb\nach\agb_\GR$ and
$\pi_\GR:\AbGb\nach\agb_\GR$ are surjective for all graphs $\GR$.
\ecorr
\bpf
Let $[h_\GR]\in\agb_\GR\ident\Ab_\GR/\Gb_\GR$. From Proposition 
\ref{satz:pi_GR=surj}
follows the existence of an $h\in\Ab$ with $\pi_\GR(h) = h_\GR$.
Then, $\bigl([\pi_{\GR'}(h)]\bigr)_{\GR'}\in\agb$ with
$\pi_\GR\bigl(([\pi_{\GR'}(h)])_{\GR'}\bigr) = [\pi_\GR(h)] = [h_\GR]$.
Analogously $\pi_\GR([h]) = [h_\GR]$ holds for $[h]:=\pi_\AbGb(h)\in\AbGb$,
whereas $\pi_\AbGb:\Ab\nach\AbGb$ is the canonical projection.
\qed
\epf
\bcorr
$\pi_\GR:\AbGb\nach\Ab_\GR/\Gb_\GR\ident\agb_\GR$ is open for all graphs $\GR$.
\ecorr
\bpf
This assertion comes from the surjectivity and the continuity of 
$\pi_\AbGb$, from the openness of $\pi_\GR:\Ab\nach\Ab_\GR$ 
and $\pi_{\Ab_\GR/\Gb_\GR}$ as well as from 
the commutativity of the following diagram:
\begin{center}
\vspace*{\CDgap}
\begin{minipage}{8cm}
\begin{diagram}[labelstyle=\scriptstyle,height=\CDhoehe,l>=3em]
\Ab         & \relax\rnachsurj^{\pi_\AbGb}             & \AbGb \\
\relax\dnachsurj^{\pi_\GR} &                           & \relax\dnachsurj^{\pi_\GR} \\ 
\Ab_\GR     & \relax\rnachsurj^{\pi_{\Ab_\GR/\Gb_\GR}} & \Ab_\GR/\Gb_\GR
\end{diagram}
\end{minipage}.
\end{center}
\qed
\epf
\label{abschn:induzhaarmasz}
Every measure on a compact $\Ab$ induces a measure on $\AbGb$ via
\bdf
Let $\mu$ be a Borel measure on $\Ab$. 

Define $\mu_\Gb (U) := \mu(\pi_{\AbGb}^{-1}(U))$ for all Borel sets 
$U$ on $\AbGb$.
\edf
\bprop
$\mu_\Gb$ is a Borel measure on $\AbGb$ for all Borel measures $\mu$
on $\Ab$.
\eprop
Especially, the induced Haar measure can be transferred from $\Ab$ to
$\AbGb$.

\section{Denseness Lemma for Projective Limits}
\label{app:lem_dense_projlim}
\blem
Let $A$ be a set, $X_a$ be a topological space for each $a\in A$ and 
$\leq$ be a partial ordering on $A$. Let
$\pi_{a_1}^{a_2}:X_{a_2}\nach X_{a_1}$ for all $a_1\leq a_2$ be a 
continuous and surjective map with 
$\pi_{a_1}^{a_2}\circ\pi_{a_2}^{a_3} = \pi_{a_1}^{a_3}$ if
$a_1\leq a_2\leq a_3$.
Furthermore, let $\pi_a : \varprojlim_{a'\in A} X_{a'} \nach X_a$ be
the usual projection on the $a$-component and $X$ be some subset of
$\varprojlim_{a\in A} X_{a}$.

Then $X$ is dense in $\varprojlim_{a\in A} X_{a}$ if
\bnum{2}
\item
$A$ is directed, i.e. for any two $a',a''\in A$ there is an $a\in A$
with $a',a''\leq a$, and
\item
$\pi_a(X)$ is dense in $X_a$ for all $a\in A$.
\enum
\elem
\bbew
Let $U\teilmenge\varprojlim_a X_a$ be open and non-empty, i.e. 
$U \obermenge \bigcap_i \pi_{a_i}^{-1}(V_i)\neq\leeremenge$ with open
$V_i\teilmenge X_{a_i}$ and finitely many $a_i\in A$.
Since $A$ is directed, there is an $a\in A$ with $a_i\leq a$ for all $i$
and thus $U \obermenge \pi_a^{-1}\bigl(\bigcap_i (\pi_{a_i}^a)^{-1}(V_i)\bigr)$ 
with non-empty $V:=\bigcap_i (\pi_{a_i}^a)^{-1}(V_i)\teilmenge X_a$.
$V$ is open because $\pi_{a_i}^a$ is continuous. 
Since $\pi_a(X)$ is dense in $X_a$ for all $a$, 
there is an $x\in X$ with $\pi_a(x)\in V$ and 
so $\pi_{a_i}(x)\in V_i$ for all $i$, hence $x\in U$.
\qed
\ebew

\addcontentsline{toc}{section}{Literaturverzeichnis}
\bibliographystyle{plain}

\begin{thebibliography}{10}

\bibitem{a72}
Abhay Ashtekar and C.~J. Isham.
\newblock Representations of the holonomy algebras of gravity and nonabelian
  gauge theories.
\newblock {\em Class. Quant. Grav.}, {\bf 9}:1433--1468, 1992.

\bibitem{a48}
Abhay Ashtekar and Jerzy Lewandowski.
\newblock Representation theory of analytic holonomy {$C^*$} algebras.
\newblock {In {\em Knots and Quantum Gravity}, edited by John C. Baez
          (Oxford Lecture Series in Mathematics and its Applications),
           Oxford University Press, Oxford, 1994.}

\bibitem{a30}
Abhay Ashtekar and Jerzy Lewandowski.
\newblock Projective techniques and functional integration for gauge theories.
\newblock {\em J. Math. Phys.}, {\bf 36}:2170--2191, 1995.

\bibitem{a13}
Abhay Ashtekar and Jerzy Lewandowski.
\newblock Quantum theory of geometry. {I}: Area operators.
\newblock {\em Class. Quant. Grav.}, {\bf 14}:A55--A82, 1997.

\bibitem{d3}
John~C. Baez and Stephen Sawin.
\newblock Functional integration on spaces of connections.
\newblock {\em J. Funct. Anal.}, {\bf 150}:1--26, 1997.

\bibitem{paper2}
Christian Fleischhack.
\newblock {Gauge Orbit Types for Generalized Connections}.
\newblock MIS-Preprint 2/2000, math-ph/0001006.

\bibitem{paper4}
Christian Fleischhack.
\newblock {Stratification of the Generalized Gauge Orbit Space}.
\newblock MIS-Preprint 4/2000, math-ph/0001008.

\bibitem{paper1}
Christian Fleischhack.
\newblock {A new type of loop independence and $SU(N)$ quantum Yang-Mills
  theory in two dimensions}.
\newblock {\em J. Math. Phys.}, {\bf 41}:76--102, 2000.

\bibitem{Kelley}
John~L. Kelley.
\newblock {\em General Topology}.
\newblock D. van Nostrand Company, Inc., Toronto, New York, London, 1955.

\bibitem{e46}
Jerzy Lewandowski and Thomas Thiemann.
\newblock Diffeomorphism invariant quantum field theories of connections in
  terms of webs.
\newblock {\em Class. Quant. Grav.}, {\bf 16}:2299--2322, 1999.

\bibitem{e8}
Alan~D. Rendall.
\newblock Comment on a paper of {A}shtekar and {I}sham.
\newblock {\em Class. Quant. Grav.}, {\bf 10}:605--608, 1993.

\end{thebibliography}

\end{document}